\documentclass[revtex4,aps,nofootinbib,showpacs]{revtex4}

\usepackage{amsmath,amsfonts,amssymb,amsthm}           
\usepackage{dsfont}                                    
\usepackage{graphicx, epsfig}                          
\usepackage{multirow}
\usepackage{verbatim}

\newcommand{\sDelta}{{\scriptstyle\Delta}}

\newcommand{\pd}{\partial}
\newcommand{\ii}{\mathrm{i}}
\newcommand{\const}{\mathrm{const}}
\newcommand{\diff}{\mathrm{d}}

\newcommand{\smallO}{\mathit{o}}
\newcommand{\bigO}{\mathit{O}}           
\DeclareMathOperator{\sgn}{sgn}          
\newcommand{\avg}[1]{\langle{#1}\rangle} 
\newcommand{\degree}{^\circ}             

\newcommand{\twoCol}[2]{ \begin{pmatrix}{#1} \\ {#2} \end{pmatrix} }

\begin{document}
\title{Peculiar seasonal effects in the neutrino day-night asymmetry}
\author{Oleg~G.~Kharlanov} \email{okharl@mail.ru}
\affiliation{Faculty of Physics, Moscow State University \\1/2 Leninskie Gory, 119991 Moscow, Russia}

\begin{abstract}
   We analyze peculiar effects in the day-night asymmetry of solar neutrinos taking place due
   to their continuous observation during the night and/or the year. Namely, we show that the
   day-night effect contains both a trivial, cumulative contribution from the whole observation
   term and a number of localized terms originating from around the midnights (during the nights)
   and the two solstices (during the year). We estimate the latter contributions using asymptotical methods
   and discuss the prospects of their isolation, i.e., magnification, by contraction of the neutrino observation term
   to small neighborhoods of the localization points. In order to complement our asymptotical
   predictions derived analytically, we also perform a full numerical analysis of a temporally-weighted
   observation of the day-night effect, including the energy spectrum of the day-night
   asymmetry and an estimation of the recoil energy distributions for the elastic-scattering
   detection channel. According to both analytical and numerical results, it turns out that
   a weighted observation is able to magnify the amplitude of the peculiar contribution
   to the day-night asymmetry even to as much as several times the cumulative term, and it looks feasible
   and appealing to perform such a weighting procedure at next-generation detectors to
   revive otherwise hidden signatures of the neutrino regeneration effect in the Earth.
\end{abstract}

   \maketitle

    \section{Introduction}\label{sec:Intro}%
    Over the past 20~years, neutrino experimental techniques have undergone considerable evolution.
    Indeed, first neutrino detection facilities attempting to observe the relevant effects at the order-of-magnitude
    level have now given way to detectors able to distinguish between neutrino flavors
    and having a qualitatively better energy resolution and lower systematic errors.
    Neutrino detectors planned to be built in the forthcoming decade,
    such as JUNO and LENA, promise good energy resolution and/or
    high event rates \cite{JUNO, JUNO_yellowbook, LENA, HyperK_intent, HyperK_potential}.
    This gives a future observer an opportunity to perform not only a determination
    of the neutrino oscillation parameters, such as the mixing angles and mass-squared differences,
    but to make cross-checks of the observations paying attention to other parameters involved.
    These include the neutrino mass hierarchy, non-standard neutrino interactions, the $\theta_{13}$ angle,
    the solar model and the model of the Earth, etc. \cite{JUNO, T2K_MassHierarchy, MassHierarchy_whitepaper,
    JUNO_SME, LENA, HyperK_potential, Smirnov2015, DayaBay_sterile, MediumBaseline_nonstandardInteractions}
    Obviously, in view of such a brilliant progress in neutrino observation techniques,
    an accompanying upgrade has to be made also in the neutrino phenomenology, suggesting new effects to `crosscheck' when upgraded
    sensitivity or resolution become available.

    In the present paper, we focus on the well-known terrestrial neutrino regeneration effect of solar neutrinos
    \cite{Carlson, BahcallKrastev, SuperK_DNA, Borexino},
    claiming that there are still things to learn and test here which, however, do not require a total revolution
    in the neutrino observation techniques. Indeed, the Sun is still one of the most prospective neutrino sources
    because of its cheapness, rather well understood characteristics and a rather steady functioning. Another distinctive
    feature of solar neutrinos is the constant movement of the neutrino `scanning ray' due to the axial and orbital motion of the Earth.
    As a result, nighttime neutrinos are able to `scan' various points inside the Earth, including its core.
    Unfortunately, we have to pay a price for being able to probe deep Earth's layers with solar neutrinos:
    in order to make any conclusions on the effects observed, we have to collect (i.e., integrate over time)
    the information gathered by the `scanning ray' over a long enough observation term.
    Such a time averaging is usually (and quite reasonably) taken into account in theoretical research devoted to the
    so-called solar neutrino day-night effect (see, e.g., \cite{BahcallKrastev, Lisi_DNA, Wei}).
    However, in most cases, a na\"ive time averaging not paying attention to quite a
    complicated solar motion around the celestial sphere leads to an estimation for the day-night asymmetry which
    depends only on the structure of the Earth just under the neutrino detector and misses other characteristics of the density
    distribution inside the Earth \cite{Wei, DNA_Kharlanov_PRD, Smirnov_perturbativeRegEffect}.
    In the present paper, we are going to revive certain observables that
    get hidden (smeared out) by the na\"ive averaging mentioned.

    \vspace{0.5em}
    The key question one has to address aiming to undertake such a revival is which types of averaging (or smearing) are technologically unavoidable in
    principle and which are not, and, if some of them can be avoided, what resolution can be achieved for the corresponding
    individual (unaveraged) quantities. Typically, there are two types of averaging in solar neutrino experiments, the averaging over the
    neutrino spectrum (i.e., over the energy $E$) and the averaging over the observation term (i.e., over the time $t$).

    Firstly, the energy resolution of neutrino detectors is indeed quite poor yet in the solar neutrino energy band
    ($\text{few}~100\text{'s~keV} < E < 20~\text{MeV}$) \cite{SuperK, SuperK_DNA_latest, Borexino_futurePlans}.
    Certain progress is going to be achieved at the liquid-scintillator JUNO detector planned to begin to operate in
    2019--2020, with the \emph{electron/positron} energy resolution $\sigma_T/T \sim 3\% / \sqrt{T / 1~\text{MeV}}$ \cite{JUNO_yellowbook}, but for ${}^8$B solar neutrinos, which are to be observed in the elastic-scattering channel,
    the \emph{neutrino} energy resolution $\sigma_E/E$ will remain modest even at JUNO. In principle though,
     there are also monochromatic ${}^7$Be solar neutrinos which do not require a neutrino detector with
      a high energy resolution. Moreover, even though matter effects are known to be minuscule for ${}^7$Be neutrinos,
     one could benefit from their high monochromaticity attempting to observe these effects, namely,
     by using a specific, time-dependent data processing (see a recent implementation of this idea in Ref.~\cite{Smirnov2015}).
     Our analysis which follows can be applied to both beryllium and boron neutrinos; it turns out to be most appealing
     for $\sim10\text{-MeV}$ boron neutrinos, preferably observed at a detector with the energy resolution $\sigma_{E}$ of about $1-2\text{ MeV}$.

    Secondly, the time averaging that is usually carried out both in numerical simulations and in theoretical estimations of
    the day-night effect is, in fact, an artificial thing that is not performed at all at detectors observing neutrinos.
    Indeed, the experiments observe individual neutrino events with perfect time stamps;
    moreover, the data processing procedure can be freely configured to select a group of events
    from the whole observation term usually lasting several years, or to temporally weight them.

    Thus, we conclude that, in order to figure out what manifestations of the terrestrial neutrino regeneration effect
    could be seen by the next-generation detectors, one should not make the averaging over the energy at first steps of the calculations,
    performing it at the end of them if necessary, and also seek to benefit from the perfect `time resolution' of the neutrino events.

    \vspace{0.5em}
    In a recent paper \cite{DNA_Kharlanov_PRD}, together with co-authors,
    we have attempted to accomplish such a task, describing solar neutrinos
    with a given energy $E$ propagating through the Earth and observed during the whole year, with the integration over the neutrino energy
    performed afterwards. Although we followed an analytical approach to the problem up to the last step (the integration over $E$),
    we have still managed to deal with the realistic trajectory of annual solar motion, which depends parametrically on the
    latitude of the neutrino detector. As a result, it turned out that even for such a non-trivial trajectory,
    the time integral over the whole year can be evaluated approximately,
    revealing next-to-leading-order (NLO) corrections to the day-night effect that, in principle, could be observable at an
    energy-resolving neutrino detector. Moreover, in a slightly counterintuitive way, these corrections come
    from the small neighborhoods of the so-called stationary points (saddle points) of the time integral over the year,
    which correspond to the winter and the summer solstices \cite{StationaryPhase, Erdelyi}.
    In other words, even though the leading-order (LO)
    contribution to the day-night asymmetry is the net effect \emph{accumulated} during the whole year,
    the first NLO corrections to it are \emph{localized in time} around the two solstices.
    Note that both energy integration and time averaging suppress the corrections, whereas the LO result persists.
    Moreover, the inner structure of the Earth (deeper than several oscillation lengths under the detector)
    manifests itself only in the NLO contributions. Such an unusual behavior of the time average of the day-night asymmetry
    leads us to thinking about the utility of a smart observation (more specifically, smart data processing) technique which
    would not lose the NLO corrections but would seek to isolate and magnify them instead by taking into account
    the time stamps and/or the energies of the neutrino events.
    For example, in Ref.~\cite{Smirnov2015}, the authors have taken up a similar problem of a smart, time-dependent observation
    of the regeneration effect for ${}^7\text{Be}$ neutrinos, with an account of a small but finite ${}^7\text{Be}$ line width.
    Another implementation of the idea is the time-domain Fourier analysis of the
    solar neutrino events to be performed at the LENA detector, in search of regular time variations of the solar neutrino
    flux related to specific solar physics \cite{LENA}.

    \vspace{0.5em}
    Thus, our paper is devoted to studying the time localization effect in the day-night asymmetry mentioned above,
    to revealing its spectral and temporal signatures, and also to analyzing the utility and observability of these effects at next-generation neutrino detectors.
    Namely, in Sec.~\ref{sec:Localization}, we discuss how localized NLO terms arise in the day-night asymmetry averaged
    over the annual solar motion and make a reminder on the mathematical properties of integrals
    containing rapidly-oscillating functions which generate localized terms.
    In Sec.~\ref{sec:Analytical}, using asymptotical methods, we analytically estimate
    the corrections to the day-night asymmetry coming from deep interfaces inside the Earth (such as the
    core-mantle interface) and estimate their time localization degree for various detector latitudes. We also consider
    separately a detector placed exactly at the tropic, since, for such a case, our analytical estimation for
    the NLO contribution to the day-night asymmetry formally diverges in the leading approximation
    and one needs to take into account higher-order terms in the relevant expansions.
    Sec.~\ref{sec:NumericalSimulation} is devoted to the numerical simulation of the effects discussed in
    the previous sections within the so-called Preliminary Reference Earth Model (PREM) \cite{PREM},
    since the small parameters in the analytical asymptotic expansions we used are actually not infinitesimal.
    Nevertheless, it turns out that the analytical estimations provide quite an accurate picture of
    the effects. This, together with the prospects of the observation of the effects studied, is discussed in
    Sec.~\ref{sec:NumericalSimulation} and in the final section~\ref{sec:Conclusion}.

    \section{The neutrino day-night effect averaged over time. Time localization of the averages}\label{sec:Localization}

    In this section, we outline the main source of the effect of localization of the day-night asymmetry,
    postponing a detailed derivation of the expressions (in fact, quite complicated ones) to the next section.
    We also cite here the background results which we will use in our transformations in Sec.~\ref{sec:Analytical}.

    \vspace{0.5em}
    In the small-$\theta_{13}$ approximation, oscillations of solar neutrinos virtually occur between the initial
    flavor state $\nu_e$ and a certain combination of $\nu_\mu$ and $\nu_\tau$, which we will denote $\nu_x$ \cite{Giunti_NeutrinoAstrophysics, PDG2014}.
    Then, the two oscillation amplitudes $\nu_e \to \nu_{e,x}$ in this, two-flavor approximation obey a Schroedinger-like
    Mikheev--Smirnov--Wolfenstein (MSW) equation \cite{Wolfenstein, MikheevSmirnov}
    \begin{gather}\label{MSW_equation}
        \ii \frac{\pd \Psi(x; E)}{\pd x} = \lambda H(x;E) \Psi(x;E), \\
        H(x;E) =  \begin{pmatrix}
                    -\cos2\theta_0 + \eta(x;E) & \sin2\theta_0 \\
                    \sin2\theta_0              & \cos2\theta_0 - \eta(x;E),
                \end{pmatrix}, \quad
        \Psi(x;E) \equiv \twoCol{\mathcal{A}_{\nu_e(0) \to \nu_e(x)}}{\mathcal{A}_{\nu_e(0) \to \nu_x(x)}},
    \end{gather}
    where $\lambda \equiv \sDelta m^2 / 4E$, $\eta(x;E) = 2\sqrt{2} G_{\text{F}} E N_e(x) / \sDelta{m^2}$,
    $E$ is the neutrino energy, $x$ is the coordinate along the neutrino ray, $x = 0$
    being the creation point; $G_{\text{F}}$ is the Fermi constant, $N_e(x)$ is the electron number density
    in the point $x$. The neutrino mixing angle for solar neutrinos $\theta_0 \equiv \theta_{12} \approx 33\degree$ and
    the corresponding mass-squared difference $\sDelta{m^2} \equiv m_2^2 - m_1^2 \approx 7.6\times10^{-5}~\text{eV}^2$ \cite{PDG2014}
    describe vacuum oscillations, while the $\eta$ parameter manifests the effect of the medium of propagation on the neutrino forward scattering \cite{Wolfenstein}.
    \begin{figure}[tbh]
        \includegraphics[width=10cm]{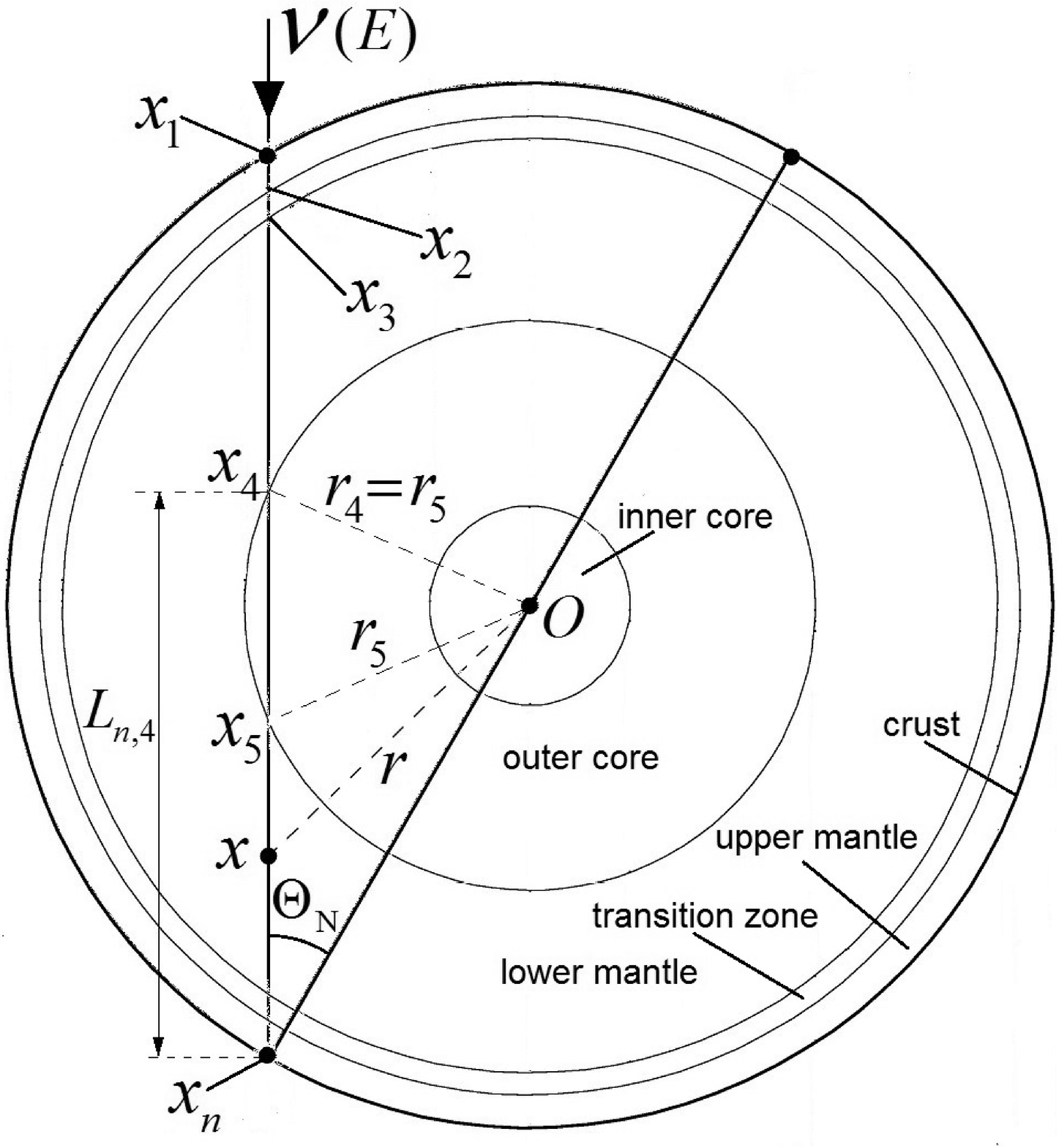}
        \caption{A nighttime neutrino passing through the Earth and the definitions of the coordinates $x_1,\ldots,x_n$,
        the distances $L_{n,j} = x_n - x_j$, the solar nadir angle $\Theta_{\text{N}}$, and the radii $r_j$ of the spherical layers inside the Earth.
        Note that the spherical layers in the figure are presented only schematically, and the structure of the PREM
        density profile is more complicated \cite{PREM}}
        \label{fig:NuRay_Geometry}
    \end{figure}

    The equation of the above type admits the so-called adiabatic approximation \cite{MikheevSmirnov}
    based on the assumption that the density of the medium $N_e(x)$ does not change rapidly along the neutrino ray, namely, that the density change
    is small within the distance of one oscillation length $\ell_{\text{osc}} = \pi / \lambda$.
    Further, it turns out that the oscillations of typical solar neutrinos are highly
    adiabatic inside the Sun \cite{MikheevSmirnov, Olivo_adiabatic, Giunti_NeutrinoAstrophysics}. Moreover, within the so-called PREM model
    \cite{PREM}, the Earth's density profile contains a number of layers,
    in each of which the propagation is also adiabatic, while
    the interfaces between them are thin compared with the oscillation length.
    Then, by using the adiabatic solution of \eqref{MSW_equation} in each adiabaticity segment,
    one arrives at the oscillation probabilities for solar neutrinos (see, e.g., \cite{DNA_Kharlanov_PRD, Wei})
    \begin{eqnarray}\label{valleyCliff}
        P_{\nu_e}(\Theta_{\text{N}}; E) &\approx& \begin{cases}
                                                    \frac12 +  \frac12 \cos2\theta_{\text{Sun}}\cos2\theta_0, & \Theta_{\text{N}} > \pi/2 \text{ (day)}, \\
                                                    \frac12 +  \frac12 \cos2\theta_{\text{Sun}}
                                                    \bigl\{\cos2\theta_n^- + 2\sin2\theta_0\sum\limits_{j=1}^{n-1}
                                                        \sDelta\theta_j\cos2\sDelta\psi_{n,j}
                                                    \bigr\},
                                                    &
                                                    \Theta_{\text{N}} < \pi/2 \text{ (night)},
                                                \end{cases}
        \\
        P_{\nu_x}(\Theta_{\text{N}}; E) &\approx& 1 - P_{\nu_e}(\Theta_{\text{N}}; E).
    \end{eqnarray}
    In the above expression, $\Theta_{\text{N}}$ is the nadir angle \cite{Astronomy} and a nighttime ($0 \le \Theta_{\text{N}} < \pi/2$) neutrino
    is assumed to cross the interfaces between the Earth's spherical layers in the points $x_1,\ldots,x_n$, \;
    $x_1$ and $x_n$ being the entry point into the Earth and the point of the detector, respectively (see Fig.~\ref{fig:NuRay_Geometry}).
    The effective mixing angles \cite{MikheevSmirnov} in the medium are defined as
    \begin{gather}
        \omega(x;E)\sin 2\theta(x) = \sin2\theta_0, \quad \omega(x;E) \cos2\theta(x) = \cos2\theta_0 - \eta(x;E), \quad \theta(x) \in [0, \pi/2],\\
        \omega(x;E) = \sqrt{1 - 2 \eta(x;E) \cos2\theta_0 + \eta^2(x;E)},     \label{omega_def}
    \end{gather}
    so that $\theta_{\text{Sun}}$ is the effective mixing angle in the solar core where the neutrino was created, while
    $\theta_j^\pm \equiv \theta(x_j \pm 0)$, $\sDelta\theta_j \equiv \theta(x_j+0) - \theta(x_j - 0)$. In particular,
    the angle $\theta_n^-$ entering Eq.~\eqref{valleyCliff} is determined by the rock density in the Earth's crust immediately under the
    neutrino detector. Finally, the oscillation phase incursion $\sDelta\psi_{n,j}$ between the $j$th crossing point ($x_j$)
    and the detector ($x_n$) is
    \begin{equation}\label{deltaPsi_nj_def}
        \sDelta\psi_{n,j} = \lambda \int\limits_{x_j}^{x_n} \omega(x;E)\diff{x}.
    \end{equation}
    It is also worth mentioning that numerical simulation justifies a high accuracy of approximation \eqref{valleyCliff}
    for both ${}^7$Be neutrinos ($E = 0.862\text{ MeV}$) and ${}^8$B solar neutrinos ($E \sim 4-12\text{ MeV}$) \cite{DNA_Kharlanov_PRD, Wei}.

    \vspace{0.5em}

    Note now that the phase incursions \eqref{deltaPsi_nj_def} are functions of the solar nadir angle $\Theta_{\text{N}}$ and the neutrino energy $E$,
    moreover, the time dependence of the nadir angle $\Theta_{\text{N}}(t)$ during the year is quite complicated \cite{Astronomy}.
    At the same time, it was shown in Ref.~\cite{DNA_Kharlanov_PRD} that despite this complexity, one can find approximate
    expressions for the daytime/nighttime electron neutrino oscillation probabilities averaged over a year-long observation
    \begin{equation}\label{yearAvg_def}
        \langle P_{\nu_e}(\,^{\text{night}}_{\text{day}};E) \rangle_{\text{year}} \equiv \int\limits_{\text{1~year}}\frac{\diff{t}}{0.5\text{~year}}\;
                                                                    \vartheta(\pm \pi/2 \mp \Theta_{\text{N}}(t)) \; P_{\nu_e}(\Theta_{\text{N}}(t); E),
        \qquad \vartheta(x) \equiv \begin{cases}
                                    1, & x \ge 0, \\
                                    0, & x < 0,
                              \end{cases}
    \end{equation}
    in closed form, using purely analytical asymptotical methods (note that $0.5~\text{year}$ in the denominator above
    is the total duration of all nights/days over the whole year). Without quoting the very form of the resulting expression yet
    (see Sec.~\ref{sec:Analytical} for details), let us describe now why it turned out to be feasible
    to find the time average of $P_{\nu_e}(\Theta_{\text{N}}(t);E)$, because the reason for that lies in the very basis of the present paper.

    \vspace{0.5em}
    Namely, analytical evaluation of \eqref{yearAvg_def} is based on the fact that
    the phase incursions $\sDelta\psi_{n,j}(\Theta_{\text{N}}(t); E)$ for the interfaces corresponding to
    the most definitive electron density jumps inside the Earth (primarily, the core-mantle interface)
    are much greater than $\pi$ and, as the Sun ascends and descends, the cosines
    $\cos 2\sDelta\psi_{n,j}(\Theta_{\text{N}}(t); E)$ entering the probabilities~\eqref{valleyCliff}
    oscillate rapidly as functions of time. Rapidness of the oscillations
    is formally controlled by a large parameter~$\lambda$ (see Eq.~\eqref{deltaPsi_nj_def}). Fortunately,
    there exists a mathematical tool for evaluating integrals of such rapidly oscillating functions in the short-oscillation-length
    limit $\lambda \to +\infty$, known as the stationary phase (saddle point) approximation \cite{StationaryPhase, Erdelyi}.
    It states that for two real-valued functions $f(t)$ and $S(t)$ that are smooth on a segment $[a,b]$ containing $n_s \ge 0$
    isolated non-degenerate stationary points $t_p \in (a,b)$ such that $S'(t_p) = 0$, $S''(t_p) \ne 0$, an asymptotic expansion
    is valid
    \begin{eqnarray}
        \int\limits_a^b f(t) \cos{\lambda S(t)} \;\diff{t} &=& \sum\limits_{p=1}^{n_s}
        \sqrt{\frac{2\pi}{|\lambda S''(t_p)|}} f(t_p) \cos\Bigl\{\lambda S(t_p) + \frac{\pi}{4} \sgn{S''(t_p)} \Bigr\}
        \nonumber\\
        &+& \left.\frac{f(\tau)\sin{\lambda S(\tau)}}{\lambda S'(\tau)}\right|_a^b + \bigO(\lambda^{-3/2}),
        \qquad \lambda \to +\infty.
        \label{statPhase}
    \end{eqnarray}
    Indeed, all rapid oscillations of the cosine almost cancel each other on integration, except for the moments of time where these oscillations
    `freeze', i.e., where the phase of the rapid oscillations $\lambda S(t)$ encounters stationary points $S'(t) = 0$.
    In the case of nighttime solar neutrinos \eqref{valleyCliff}, the role of the function $\lambda S(t)$ is played by
    $2\sDelta\psi_{n,j}(\Theta_{\text{N}}(t); E)$, while $f(t)$ are constants proportional to $\sDelta\theta_j$.
    Then, the desired $\lambda\to +\infty$ asymptotic of the year-average day-night asymmetry
    $\langle P_{\nu_e}(\text{night};E) \rangle_{\text{year}} - \langle P_{\nu_e}(\text{day}; E) \rangle_{\text{year}} $
    is obtained by applying \eqref{statPhase} to the adiabatic probabilities \eqref{valleyCliff}
    two times: first to a time integral over a given night and after that to a sum over 365 nights
    which is also virtually an integral \cite{DNA_Kharlanov_PRD}. The result contains a trivial
    non-oscillating term $\frac12 \cos2\theta_{\text{Sun}} (\cos2\theta_n^- - \cos2\theta_0)$,
    plus the contributions of two stationary points resulting from applying the approximation
    \eqref{statPhase} to the sum ($\sim$ integral) over 365 nights. These two   points
    are the winter and the summer solstices \cite{DNA_Kharlanov_PRD}.

    \vspace{0.5em}
    A key property of the time average of the day-night asymmetry that was not studied in detail in
    \cite{DNA_Kharlanov_PRD} and to which we are going to draw major attention in the present paper,
    is the localization of contributions to integrals of rapidly oscillating functions. Namely,
    note that the leading contribution to such an integral \eqref{statPhase} depends only on
    the values of the functions $f(t)$ and $S(t)$ and their derivatives in the stationary points $t = t_p$ rather than on
    the whole interval $[a,b]$. Moreover, this contribution is generated within several
    periods of the function $\cos \lambda S(t)$ around $t = t_p$, where one can use
    the Taylor expansion $S(t) \approx S^{(2)}(t) \equiv S(t_p) + \frac12 S''(t_p) (t-t_p)^2$,
    $f(t) \approx f(t_p)$,
    \begin{eqnarray}
        \int\limits_{t_p - \sDelta{t}/2}^{t_p + \sDelta{t}/2} f(t)\cos{\lambda S(t)}\; \diff{t} &=&
        \int\limits_{-\infty}^{+\infty} f(t_p)\cos\{\lambda S^{(2)}(t)\}\; \diff{t} + \bigO(1/\lambda), \qquad \lambda \to +\infty,
        \label{localization_example}\\
        \int\limits_{-\infty}^{+\infty} f(t_p)\cos\{\lambda S^{(2)}(t)\}\; \diff{t} &=& \sqrt{\frac{2\pi}{|\lambda S''(t_p)|}}
        f(t_p) \cos\Bigl\{\lambda S(t_p) + \frac{\pi}{4} \sgn{S''(t_p)} \Bigr\}, \label{localization_leadingAsymp}
    \end{eqnarray}
    where $\sDelta{t}$ is a small enough fixed positive number. The subsegments between the stationary points where $|S'(t)| \ge \const > 0$ contribute to
    the integral \eqref{statPhase} as $\smallO(\lambda^{-N})$ for any positive $N$ \cite{StationaryPhase, Erdelyi}.

    \vspace{0.5em}
    The above localization property leads us to the following trick that will be implemented in what follows:
    if one defines the time average of $f(t) \cos{\lambda S(t)}$ as the time integral \eqref{localization_example}
    divided by the observation time $\sDelta{t}$, then, say, a 10-times reduction of $\sDelta{t}$
    will lead to a 10-fold magnification of the stationary-point contribution to the average!
    Note that nothing like happens to time averages of non-oscillating functions, since
    the corresponding time integrals are proportional to the observation time $\sDelta{t}$
    for small $\sDelta{t}$. In connection to neutrinos, we expect that the stationary-point
    contributions to the day-night asymmetry that are localized near the two solstices should
    get amplified when one shrinks the observation period to one of these stationary points.
    Still, before we resort to the very analysis of the localization of the day-night asymmetry
    of solar neutrinos, there are a couple of general things on the localization effect
    to discuss here.

    \vspace{0.5em}
    Firstly, the time integral~\eqref{localization_example} is localized in the point $t = t_p$ only in the limit $\lambda \to +\infty$.
    For large but finite $\lambda$, the leading $\bigO(\lambda^{-1/2})$ asymptotic of \eqref{localization_example},
    as it was mentioned above, is generated within several periods of the oscillating cosine of both \eqref{localization_example}
    and \eqref{localization_leadingAsymp}, i.e. within $t - t_p = \bigO(\delta{t}_p)$, where the time localization scale
    $\delta{t}_p$ of the contribution of the $p$th stationary point is defined as
    \begin{equation}\label{locScale_def}
        |\lambda S(t_p + \delta{t}_p) - \lambda S(t_p)| = 2\pi, \quad \delta{t_p} > 0, \qquad \lambda \to +\infty,
    \end{equation}
    i.e., in our case $S'(t_p) = 0$, $S''(t_p) \ne 0$,
    \begin{equation}\label{locScale_quadratic}
        \delta{t_p} \approx \sqrt{\frac{4\pi}{|\lambda S''(t_p)|}}, \qquad \lambda \to +\infty.
    \end{equation}
    In the case of degenerate stationary points, e.g., if $S'(t_p) = S''(t_p) = S'''(t_p) = 0$, $S''''(t_p) \ne 0$,
    the localization also takes place \cite{StationaryPhase}, but the definition \eqref{locScale_def} leads to
    \begin{equation}\label{locScale_quartic}
        \delta{t_p} \approx \left[\frac{48\pi}{|\lambda S''''(t_p)|}\right]^{1/4}, \qquad \lambda \to +\infty.
    \end{equation}
    Note that in principle, the expression \eqref{locScale_quadratic} for the localization scale in the quadratic approximation
    is accurate for large values of $\lambda$, namely, for $\lambda \gg |S'''(t_p)|^2 / |S''(t_p)|^3, |S''''(t_p)| / |S''(t_p)|^2$, etc.,
    i.e., it is valid when the quadratic approximation $S(t)\approx S^{(2)}(t)$ is accurate within the localization domain around $t = t_p$.
   Analogously, Eq.~\eqref{locScale_quartic} is valid for such large $\lambda$
   that the approximation $S(t) \approx S(t_p) + S''''(t_p) (t-t_p)^4 / 24$
   is accurate within $|t-t_p| \lesssim \delta{t_p}$.

    \vspace{0.5em}
    Secondly, in order to make the contribution of the stationary point more vivid in Eq.~\eqref{statPhase} and avoid the boundary term,
    it is worth using a weighted time integral $\int w(t) f(t) \cos{\lambda S(t)} \, \diff{t}$  instead of a definite intergral
     $\int_{t_p - \sDelta{t}/2}^{t_p + \sDelta{t}/2} f(t) \cos{\lambda S(t)} \, \diff{t}$,
    with the weighting function $w(t)$ smoothly tending to zero at the ends of the interval of integration. The
    shrinkage of the interval of averaging mentioned above is then achieved by choosing the weighting function more and more
    concentrated around the desired stationary point. In connection to the neutrino day-night asymmetry,
    such an approach implies introduction of temporally weighted averages instead of \eqref{yearAvg_def}
    \begin{eqnarray}\label{wAvg_def}
        \langle P_{\nu_e}(\,^{\text{night}}_{\text{day}};E) \rangle_{w} &\equiv& \int\limits_{\text{1~year}}\frac{w(t)\diff{t}}{T_{\text{night,day}}}\;
                                                                    \vartheta(\pm \pi/2 \mp \Theta_{\text{N}}(t)) \; P_{\nu_e}(\Theta_{\text{N}}(t); E), \\
        T_{\text{night,day}} & \equiv & \int\limits_{\text{1~year}}w(t)\diff{t} \; \vartheta(\pm \pi/2 \mp \Theta_{\text{N}}(t)),
        \label{T_nightday}
    \end{eqnarray}
    where the weighting function $w(t)$ is assumed to obey the normalization condition
    \begin{equation}\label{w_normalization}
        \int\limits_{\text{1~year}}\frac{w(t)\diff{t}}{1\text{~year}} = 1
    \end{equation}
    and $T_{\text{night,day}}$ are the weighted total nighttime and daytime, respectively.
    In order to legally apply the stationary phase approximation \eqref{statPhase} to the weighted averages \eqref{wAvg_def},
    we should use  weighting functions $w(t)$ that are smooth functions of the season $\varsigma = \varsigma(t) \in [0, 2\pi)$
    and the time of day $\tau = \tau(t) \in [0, 2\pi)$.
    In terms of the averages thus defined, one can describe solar neutrino observations made during certain seasons or hours
    and study the properties of the contributions localized near the solstices.

    \vspace{0.5em}
    Finally, let us pay attention to the fact that a neutrino experiment does not measure the probabilities
    $P_{\nu_{e,x}}$ directly but rather counts the neutrino events. As a consequence, every average it measures is
    accompanied by a nonzero uncertainty, and limitation of the observation period obviously enhances
    the statistical uncertainty. Nevertheless, we claim that the signal-to-noise ratio for time-localized contributions
    to the day-night asymmetry is improved by this limitation. To demonstrate that, let us consider a Poisson-distributed time series
    of neutrino events observed at times $t_1, t_2, \ldots, t_{N_{\text{obs}}}$ during 1~year and define
    the weighted total numbers of nighttime/daytime events
    \begin{equation}\label{N_w}
        N_{\text{night,day}}^{(w)} = \sum\limits_{k=1}^{N_{\text{obs}}} \vartheta(\pm \pi/2 \mp \Theta_{\text{N}}(t_k)) \;w(t_k).
    \end{equation}
    These event numbers have the following expectation values and variances
    \begin{eqnarray}\label{Nnight}
        \mathds{E}[N_{\text{night,day}}^{(w)}] &=&  \int\limits_{1~\text{year}} \frac{\vartheta(\pm\pi/2 \mp \Theta_{\text{N}}(t)) w(t)\diff{t}}{0.5\text{~year}}\int \Phi(E)\diff{E}\;
                            \bigl\{ \sigma_{\nu_e}(E) P_{\nu_e}(\Theta_{\text{N}}(t); E) + \sigma_{\nu_x}(E) P_{\nu_x}(\Theta_{\text{N}}(t); E) \bigr\}
        \nonumber\\
        &=&  \frac{T_{\text{night,day}}} {0.5\text{~year}} \int \Phi(E)\diff{E}\;
                            \bigl\{ \sigma_{\nu_e}(E) \avg{P_{\nu_e}(\,^{\text{night}}_{\text{day}}; E)}_w
                            + \sigma_{\nu_x}(E) \avg{P_{\nu_x}(\,^{\text{night}}_{\text{day}}; E)}_w \bigr\},
                            \\
        \label{Nnight_variance}
        \mathds{V}[N_{\text{night,day}}^{(w)}] &=&
                            \int\limits_{1~\text{year}} \frac{\vartheta(\pm\pi/2 \mp \Theta_{\text{N}}(t)) w^2(t)\diff{t}}{0.5\text{~year}}\int \Phi(E)\diff{E}\;
                            \bigl\{ \sigma_{\nu_e}(E) P_{\nu_e}(\Theta_{\text{N}}(t); E) + \sigma_{\nu_x}(E) P_{\nu_x}(\Theta_{\text{N}}(t); E) \bigr\},
    \end{eqnarray}
    where $\sigma_{\nu_{e,x}}(E)$ represent the sensitivities of the detector to $\nu_e$ and $\nu_{x}$,
    the energy bin(s) extracted from the recoil electron spectrum, and the fiducial volume of the detector, while $\Phi(E)$ is the spectrum
    of the solar neutrino flux ($\nu_e$ and $\nu_x$ together). Let us define the (effective, weighted)
    observation period $T_{\text{obs}}$ by an approximate identity $w^2(t) \simeq (1~\text{year} / T_{\text{obs}}) w(t)$, so that
    \begin{equation}\label{Nnight_variance_estimation}
        \mathds{V}[N_{\text{night,day}}^{(w)}] \simeq \frac{1~\text{year}}{T_{\text{obs}}} \; \mathds{E}[N_{\text{night,day}}^{(w)}],
    \end{equation}
    and note that, according to the definition \eqref{T_nightday}, $T_{\text{night}} + T_{\text{day}} = 1~\text{year}$,
    so both $T_{\text{night}}$ and $T_{\text{day}}$ remain finite when one shrinks the observation term $T_{\text{obs}}$.
    Within the leading approximation, the \emph{signals} $\mathds{E}[N_{\text{day}}^{(w)}]$ and $\mathds{E}[N_{\text{night}}^{(w)}]$ also remain finite,
    because both the daytime probabilities $P_{\nu_{e,x}}(\Theta_{\text{N}}; E)$ and the leading contribution
    to the nighttime ones \eqref{valleyCliff} are non-oscillating and are thus unaffected by time averaging.
    Therefore, according to \eqref{Nnight_variance_estimation}, the statistical uncertainty (\emph{noise}) of $N_{\text{day}}^{(w)}$
    can be made negligibly small by extending the neutrino observation period as much as possible. On the other hand,
    the relative uncertainty (\emph{noise}) of $N_{\text{night}}^{(w)}$ is
    \begin{equation}
        \frac{\mathds{V}^{1/2}[ N_{\text{night}}^{(w)} ] }{\mathds{E}[ N_{\text{night}}^{(w)}]}
        \simeq \sqrt{\frac{1~\text{year}}{T_{\text{obs}}}} \; \mathds{E}^{-1/2}[ N_{\text{night}}^{(w)}]
        \approx \sqrt{\frac{1~\text{year}}{T_{\text{obs}}}} \; \mathds{E}^{-1/2}[ N_{\text{day}}^{(w)}],
    \end{equation}
    which is approximately proportional to the inverse square root
    $1 / \sqrt{T_{\text{obs}}}$ of the observation time, whereas restriction
    of the observation term to a small neighborhood of the localization points
    of the oscillating contributions to the averages $\avg{P_{\nu_{e,x}}(\text{night}; E)}_w$
    is able to magnify these contributions (the \emph{signal}) proportionally to $1/T_{\text{obs}}$.
    The signal-to-noise ratio is thus improved proportionally to $1/\sqrt{T_{\text{obs}}}$, and
    we qualitatively conclude that a restricted observation period helps resolve the time-localized contributions
    to the day-night effect.

    \section{Seasonal effects in the day-night asymmetry: analytical estimations}\label{sec:Analytical}
    Let us study the properties of the nighttime neutrino flavor probability $P_{\nu_e}(\Theta_{\text{N}}(t); E)$ $(\Theta_{\text{N}}(t) < \pi/2)$
    weighted using a function $w(t)$, as defined in Eqs.~\eqref{wAvg_def}, \eqref{T_nightday}.
    As mentioned in the previous section, we assume that the weighting function is a smooth function
    of the season $\varsigma$ and the time of day $\tau$, \quad $w=w(\varsigma(t), \tau(t))$.

    First of all, following the approach of paper \cite{DNA_Kharlanov_PRD}, we note that due to the Heaviside theta function, the integral in Eq.~\eqref{wAvg_def}
    is in fact a sum of $N_{\text{days}} = 365$ integrals over all the nights of the year (assuming that we are not in the polar latitudes,
    so that there are no polar nights or days), namely,
    \begin{eqnarray}
        \langle P_{\nu_e}(\text{night}; E)\rangle_w &\approx& P_{\nu_e}^{\const}(\text{night};E)  +
        \frac{1}{T_\text{night}} \sum\limits_{d = 0}^{N_{\text{days}}-1} I_d,     \label{wAvg_to_I_d}\\
        P_{\nu_e}^{\const}(\text{night};E) &\equiv& P_{\nu_e}(\text{day}; E) +  \frac12 \cos2\theta_{\text{Sun}} (\cos2\theta_n^- -\cos2\theta_0), \\
        I_d &\equiv& \int\limits_{d\text{th night}} \{ P_{\nu_e}(\Theta_{\text{N}}(t); E) - P_{\nu_e}^{\const}(E) \} \; w(\varsigma(t), \tau(t)) \diff{t},
        \label{I_d}
    \end{eqnarray}
    where we have singled out the constant contribution $P_{\nu_e}^{\const}(\text{night};E)$ to the nighttime probability which does not
    require special time averaging. Let us now apply the stationary phase approximation \eqref{statPhase} to the integral $I_d$ over the night
    and calculate the relevant localization scales.

    \vspace{0.5em}
    In a realistic approximation $N_{\text{days}} \gg 1$, we can neglect the variation of the seasonal variable $\varsigma(t)$
    within a single night and adopt a reparametrization $\diff{t} = \frac{1~\text{year}}{N_{\text{days}}} \, \frac{\diff\tau}{2\pi}$.
    Next, we use the solar trajectory \cite{Astronomy}
    \begin{equation}
        \cos\Theta_{\text{N}}(\varsigma, \tau) = \cos\chi \sin\varsigma \sin\tau + \cos\varepsilon \cos\chi \cos\varsigma \cos\tau +
                                                 \sin\varepsilon \sin\chi \cos\varsigma,
    \end{equation}
    where $\varepsilon = 23.4\degree$ is the axial tilt of the Earth and $\chi$ is the latitude of the detector.
    In what follows, we assume that the latter lies between the tropic and the polar circle, i.e., that $\varepsilon < \chi < \pi / 2 - \varepsilon$;
    the results for detectors in the southern hemisphere ($-\varepsilon > \chi > -\pi / 2 + \varepsilon$)
    are clearly equivalent to those for the opposite ones in the northern hemisphere, up to a change $\text{winter} \leftrightarrow \text{summer}$.
    The phase incursions $\sDelta\psi_{n,j}(\Theta_{\text{N}}(\varsigma, \tau); E)$ entering the oscillating function
    $P_{\nu_e}(\Theta_{\text{N}}(t); E) - P_{\nu_e}^{\const}(E)$ in the integral \eqref{I_d} (see expression~\eqref{valleyCliff})
    achieve extremum values at midnight, when
    \begin{equation}\label{cosThetaN_midnight}
        \cos\Theta_{\text{N}}(\varsigma, \tau) \to \max\limits_{\tau\in [0,2\pi)} \equiv \cos\Theta_{\text{N}}^{\text{midnight}}(\varsigma)
                                               = \cos\chi \sqrt{\sin^2\varsigma + \cos^2\varepsilon \cos^2\varsigma} +
                                                 \sin\varepsilon \sin\chi \cos\varsigma,
    \end{equation}
    in other words, the midnight is the only stationary point of the integral $I_d$ over a single night. The corresponding
    localization scale \eqref{locScale_quadratic} of the contribution of the $j$th interface to $I_d$ is
    \begin{equation}
        \delta{t}_{j}^{\text{midnight}}(\varsigma) = \frac{24~\text{hrs}}{2\pi} \delta{\tau}_{j}^{\text{midnight}}(\varsigma)
        = \frac{24~\text{hrs}}{2\pi} \times
        \left.\sqrt{\frac{4\pi}{|\pd_\tau^2 (2\sDelta\psi_{n,j}(\Theta_{\text{N}}(\varsigma,\tau),E))|}}\right|_{\tau=\tau_{\text{midnight}}(\varsigma)},
    \end{equation}
    where $\tau_{\text{midnight}}(\varsigma$) is the value of $\tau$ corresponding to the midnight \eqref{cosThetaN_midnight}.
    In order to evaluate the second derivative, let us introduce the distance between the $j$th crossing point and the detector
    (see Fig.~\ref{fig:NuRay_Geometry})
    \begin{eqnarray}\label{L_nj}
        L_{n,j}(\Theta_{\text{N}}) &\equiv& x_n - x_j = r_n\cos\Theta_{\text{N}} + s' \sqrt{r_j^2 - r_n^2
        \sin^2\Theta_{\text{N}}}, \qquad  r_j > r_n \sin \Theta_{\text{N}}, \\
        s' &\equiv& \sgn\{L_{n,j} - r_n \cos \Theta_{\text{N}}\},
    \end{eqnarray}
    where $r_n \equiv r_{\text{Earth}} = 6371~\text{km}$ is the radius of the Earth and $r_j$ is the radius of the $j$th interface;
    the sign $s' = +1$ if the neutrino is leaving a more shallow Earth's layer and entering a deeper one, and $s' = -1$
    if the neutrino is going outside, leaving the deeper layer. Note that the inequality in \eqref{L_nj} requires
    that the neutrino actually cross the interface with $r = r_j$.

    For solar neutrinos that are well under the MSW resonance inside the Earth and $\eta \ll 1$ (see Eq.~\eqref{MSW_equation}),
    the definition of the phase incursions \eqref{deltaPsi_nj_def} implies $\sDelta\psi_{n,j} \simeq \lambda L_{n,j}$.
    Hence, at midnight, when $\pd_\tau \sDelta\psi_{n,j} = 0$,
    \begin{gather}
        \pd_\tau^2 \sDelta\psi_{n,j} \approx
        \left[\pd_\tau^2(\cos\Theta_{\text{N}}) \cdot \lambda\,\frac{\pd L_{n,j}}{\pd(\cos\Theta_{\text{N}})}\right]_{\text{midnight}}
        = -\mathcal{N}(\varsigma) \cdot \frac{s' \lambda L_{n,j}(\Theta_{\text{N}}^{\text{midnight}}(\varsigma))}
                                             {\sqrt{r_j^2 / r_n^2 - \sin^2 \Theta_{\text{N}}^{\text{midnight}}(\varsigma)}}, \\
        \mathcal{N}(\varsigma) \equiv \cos\chi \sqrt{\sin^2\varsigma + \cos^2\varepsilon \cos^2\varsigma}.
    \end{gather}
    Finally, the localization scale of the midnight contribution to \eqref{I_d}
    \begin{equation}\label{delta_t_midnight}
        \delta{t}_{j}^{\text{midnight}}(\varsigma) = \frac{1}{\sqrt{\mathcal{N}(\varsigma)}}
        \frac{\Bigl( r_j^2 / r_n^2 - \sin^2 \Theta_{\text{N}}^{\text{midnight}}(\varsigma) \Bigr)^{1/4}}
             {\sqrt{2\pi \lambda L_{n,j}(\Theta_{\text{N}}^{\text{midnight}}(\varsigma))}} \times 24~\text{hrs},
        \qquad
        r_j > r_n \sin \Theta_{\text{N}}^{\text{midnight}}(\varsigma).
    \end{equation}
    \begin{figure}[tbh]
        $$
        \begin{array}{cc}
            \includegraphics[width=8.7cm]{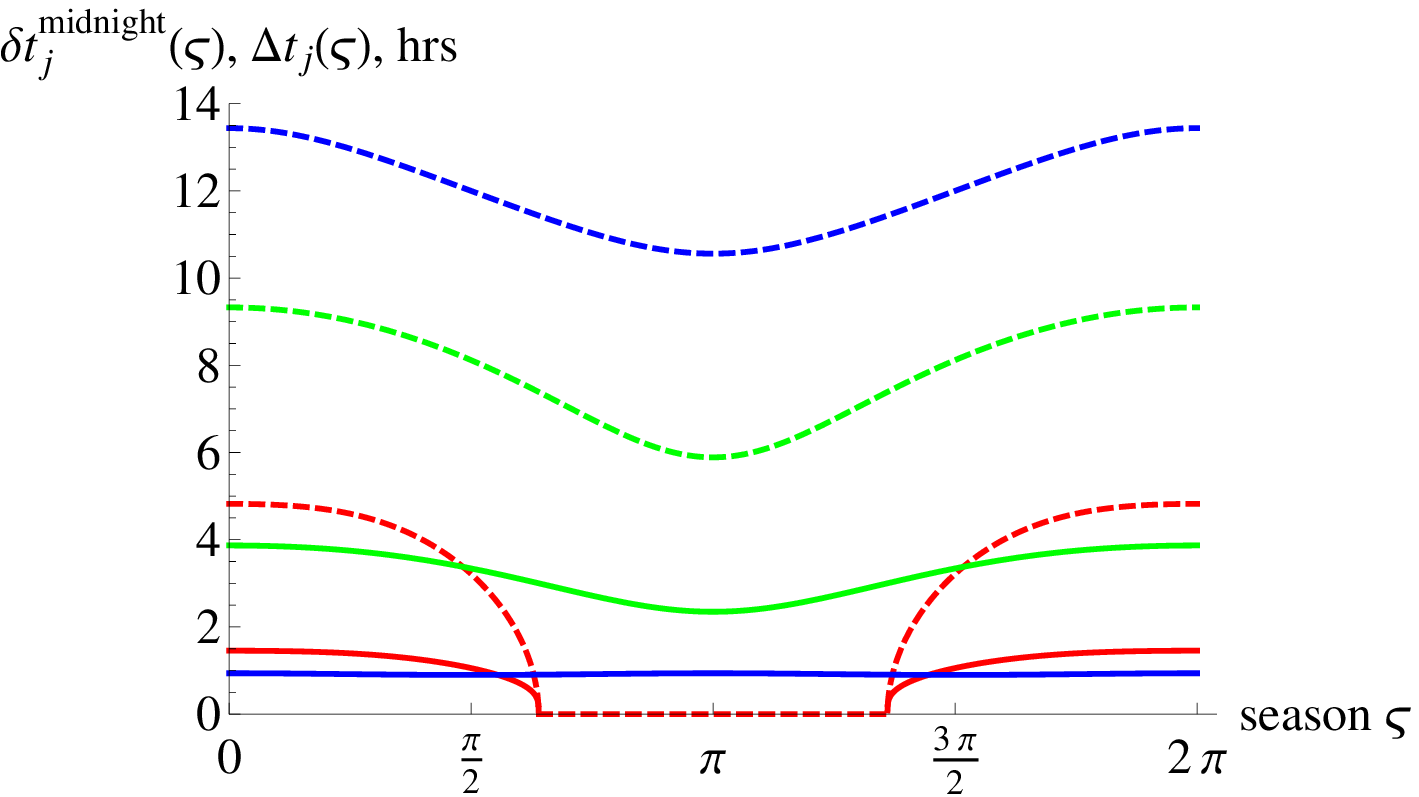}  &  \includegraphics[width=8.7cm]{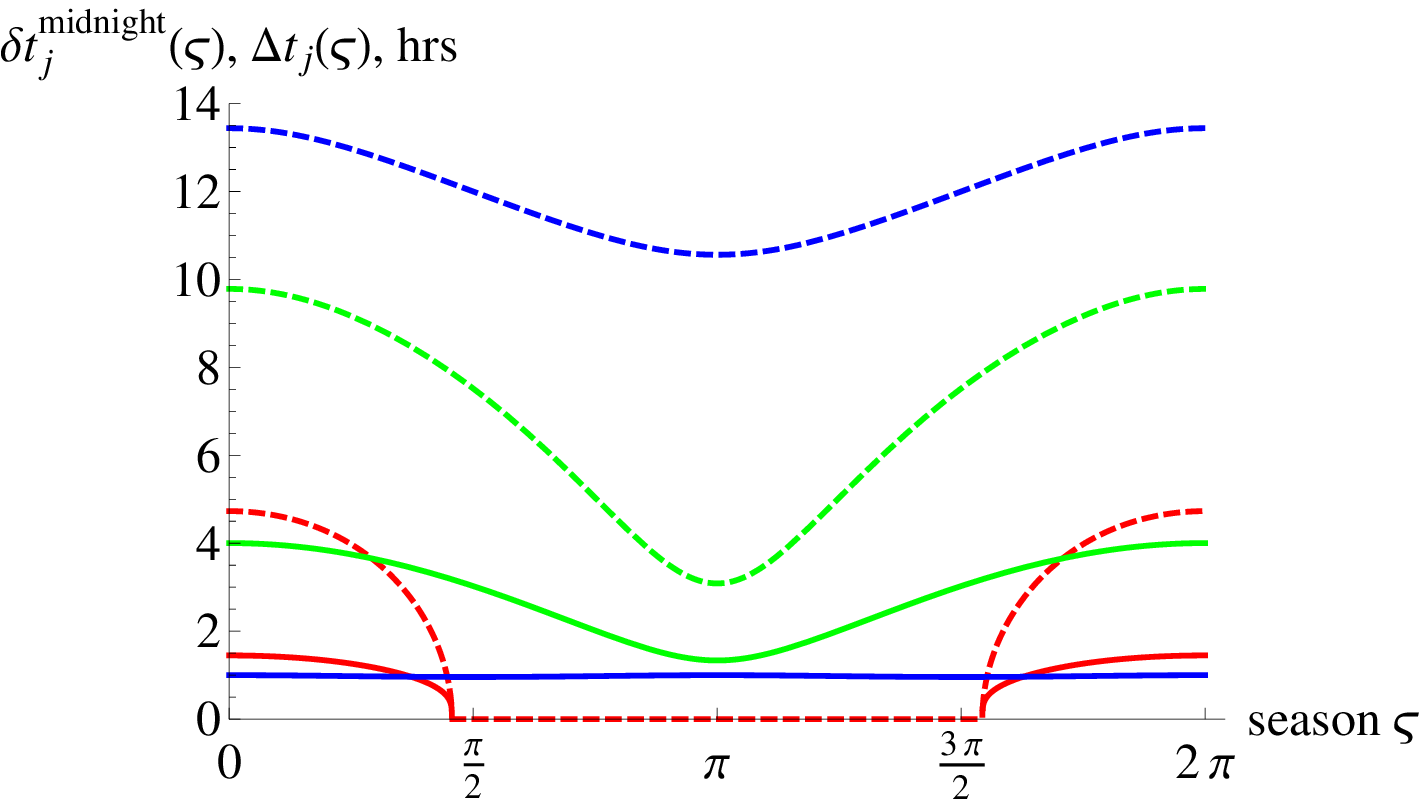} \\
            \text{(a)}                                         &  \text{(b)}
        \end{array}
        $$
        \caption{Localization scale \eqref{delta_t_midnight} for the midnight stationary-point contribution to the day-night asymmetry
        of neutrinos with $E = 10~\text{MeV}$ observed at a detector placed at the tropic $\chi = 23.4\degree$ (left) and at $\chi = 36.2\degree$ (right),
        in different seasons.
        Solid red/green/blue curves correspond to interfaces with radii $r_j = 3480~\text{km}$ (core to mantle, $s'=-1$),
        $5701~\text{km}$ (lower to upper mantle, $s'=-1$), $6371~\text{km}$ (the entry into the Earth, $s'=+1$), respectively.
        Dashed curves show the total time $\sDelta{t}_j$ during the night the Sun shines throw these interfaces (see Eq.~\eqref{delta_t_j})}
        \label{fig:delta_t_midnight}
    \end{figure}

    We have plotted the localization scale of the midnight stationary point against the season $\varsigma$
    in Fig.~\ref{fig:delta_t_midnight}, for two latitudes of the detector, $\chi = \varepsilon = 23.4\degree$
    and $\chi = 36.2\degree$, and for different interfaces, $r_j = 3480~\text{km}$ (core to mantle),
    $5701~\text{km}$ (lower to upper mantle), $6371~\text{km}$ (the entry into the Earth).
    The energy of neutrinos in the plots is $E = 10~\text{MeV}$, representing a typical ${}^8$B neutrino.
    To make the picture more informative, we have included in the plots the exposure time, i.e., total time
    over the night the neutrinos pass through the $j$th interface
    \begin{gather}\label{delta_t_j}
        \sDelta{t}_j(\varsigma) \equiv \frac{24~\text{hrs}}{2\pi} \int\limits_0^{2\pi}
                                \vartheta\left(\arcsin\frac{r_j}{r_n} - \Theta_{\text{N}}(\varsigma,\tau)\right)\;\diff\tau
                                = \frac{24~\text{hrs}}{2\pi}
                                   \times 2\mathcal{F}\left(
                                                \frac{\sqrt{1-r_j^2/r_n^2} - \sin\varepsilon \sin\chi \cos\varsigma}
                                                {\cos\chi \sqrt{\sin^2\varsigma + \cos^2\varepsilon \cos^2\varsigma}}
                                   \right),\\
        \label{arccos_modified}
        \mathcal{F}(\xi) \equiv     \begin{cases}
                                        0, & \xi > 1, \\
                                        \arccos \xi, & -1 \le \xi \le 1, \\
                                        \pi, & \xi < -1.
                                    \end{cases}
    \end{gather}
    The above quantity is nothing but a length of the interval supporting the $j$th contribution to the integrand of \eqref{I_d}, and
    the corresponding degree of localization is determined by the ratio $\sDelta{t}_j(\varsigma) / \delta{t}_j^{\text{midnight}}(\varsigma)$.
    From Fig.~\ref{fig:delta_t_midnight}, it is clear that the localization is quite high, and one is expecting considerable
    magnification of the stationary-point contributions to the day-night effect by a concentration of observations around midnights.
    We will check this effect numerically in the next section, and now let us proceed with the evaluation of the weighted average \eqref{wAvg_to_I_d}.
    \vspace{0.5em}

    Note that, as mentioned in \cite{DNA_Kharlanov_PRD}, for the integral \eqref{I_d} in question, the boundary term in the
    stationary phase approximation \eqref{statPhase} is absent, since at the boundary $\Theta_{\text{N}}(\varsigma, \tau) = \arcsin(r_j/r_n)$, the neutrino
    ray is tangent to the interface $r = r_j$ and $\pd_\tau \sDelta\psi(\Theta_{\text{N}};E) \propto \pd L_{n,j} / \pd(\cos\Theta_{\text{N}}) \to \infty$.
    On the other hand, the stationary-point term in \eqref{statPhase} is obtained straightforwardly using the localization scale just found,
    so that the stationary phase approximation for the time integral \eqref{I_d} over the $d$th night reads
    \begin{eqnarray}\label{I_d_result}
        I_d                 &\approx& \frac{0.5~\text{year}}{N_{\text{days}}} w(\varsigma, \tau_{\text{midnight}}(\varsigma)) \times
                                \cos2\theta_{\text{Sun}} \sin2\theta_0
                                \sum\limits_{j=1}^{n-1} \sDelta\theta_j
                                    \vartheta\big(r_j - r_n\sin\Theta_{\text{N}}^{\text{midnight}}(\varsigma)\big) \nonumber\\
                                    &\times&
                                    \frac{(r_j^2/r_n^2 - \sin^2\Theta_{\text{N}}^{\text{midnight}}(\varsigma))^{1/4}}
                                         {\sqrt{\pi\lambda L_{n,j}(\Theta_{\text{N}}^{\text{midnight}}(\varsigma))\,\cdot\, \mathcal{N}(\varsigma)}}
                                    \cos\bigl\{2\sDelta\psi_{n,j}(\Theta_{\text{N}}^{\text{midnight}}(\varsigma);E) - s'\frac{\pi}{4}\bigr\},
        \qquad
        \varsigma = \frac{2\pi d}{N_{\text{days}}}.
    \end{eqnarray}
    It is no surprise that the value of the integral, up to the weighting factor $w(\varsigma, \tau_{\text{midnight}}(\varsigma))$,
    coincides with the one calculated in \cite{DNA_Kharlanov_PRD} without weighting. This, together with the fact that $I_d$
    is determined by the midnight values of the time-dependent parameters involved, is a
    manifestation of the time localization discussed in Sec.~\ref{sec:Localization}.

    \vspace{0.5em}
    After having evaluated the integrals $I_d$ over all the nights in the year, we should add them together.
    Within the same approximation $N_{\text{days}} \gg 1$, this sum can be replaced by an integral,
    \begin{equation}\label{sum365toIntegral}
        \sum\limits_{d = 0}^{N_{\text{days}}-1} I_d \approx
                            \frac{N_{\text{days}}}{2\pi} \int\limits_0^{2\pi} I(\varsigma) \diff\varsigma,
        \qquad
        I_d \equiv I\left( \frac{2\pi d}{N_{\text{days}}}\right).
    \end{equation}
    Now that we have replaced the day number $d$ with the continuous seasonal variable $\varsigma$, the above integral also takes the form of
    a sum of $n-1$ integrals containing rapidly oscillating functions of the season
    $\cos\bigl\{2\sDelta\psi_{n,j}(\Theta_{\text{N}}^{\text{midnight}}(\varsigma);E) - s'\frac{\pi}{4}\bigr\}$ (see Eq.~\eqref{I_d_result}),
    and we apply the approximation \eqref{statPhase} to it once again. In this case, there is no boundary and there are
    two stationary points, namely, the solstices $\varsigma = \varsigma_s = 0,\pi, \; s=\mp1$, which correspond to the lowest and the highest midnight
    solar positions over the year. The value of the nadir angle and the second derivative of the phase in these stationary points are
    \begin{gather}
        \Theta_{\text{N}}^{\text{midnight}}(\varsigma_s) = \chi + s \varepsilon, \\
        \Bigl.\pd_\varsigma^2 \bigl(2\sDelta\psi_{n,j}(\Theta_{\text{N}}^{\text{midnight}}(\varsigma); E) \bigr)\Bigr|_{\varsigma = \varsigma_s}
        = 2s \sin(\chi + s\varepsilon)\tan\varepsilon \times \frac{s' \lambda L_{n,j}(\chi + s\varepsilon)}
                                             {\sqrt{r_j^2 / r_n^2 - \sin^2(\chi + s\varepsilon) }}, \qquad
                                r_j > r_n \sin(\chi + s\varepsilon),
    \end{gather}
    where, as we noted earlier, $\chi > \varepsilon$. Hence, the localization scale of the $s$th-solstice contribution to the seasonal integral
    \eqref{sum365toIntegral} coming from the $j$th interface is
    \begin{eqnarray}
        \delta{t}_{s,j}^{\text{solstice}} = \frac{1~\text{year}}{2\pi}\delta\varsigma_{s,j} &=&
        \frac{1~\text{year}}{2\pi}
        \left.\sqrt{\frac{4\pi}
                   {\bigl|\pd_\varsigma^2 \bigl(2\sDelta\psi_{n,j}(\Theta_{\text{N}}^{\text{midnight}}(\varsigma); E) \bigr)\bigr|}
                   }
        \right|_{\varsigma=\varsigma_s}
        \nonumber\\
        &=& \frac{(r_j^2 / r_n^2 - \sin^2 (\chi + s \varepsilon))^{1/4}}
                                        {\sqrt{2\pi \lambda L_{n,j}(\chi + s \varepsilon) \cdot \sin(\chi + s \varepsilon) \tan\varepsilon}}
                            \times 1\text{ year} \qquad (\chi > \varepsilon).
        \label{locScale_sj}
    \end{eqnarray}
    On the other hand, the seasonal integral \eqref{sum365toIntegral} in the stationary phase approximation \eqref{statPhase},
    plus the constant contribution $P_{\nu_e}^{\const}(\text{night};E)$ separated in \eqref{wAvg_to_I_d},
    give us the final expression for the weighted nighttime neutrino observation probability \eqref{yearAvg_def}, which
    coincides with the non-weighted one \eqref{wAvg_def} calculated in \cite{DNA_Kharlanov_PRD} up to two weighting factors
    $w(\varsigma_{\pm1}, \tau_{\text{midnight}}(\varsigma_{\pm1}))$,
    \begin{multline}\label{deltaP_year_w}
        \langle P_{\nu_e}(\text{night}; E) \rangle_{w} \approx P_{\nu_e}(\text{day}; E) +  \frac12 \cos2\theta_{\text{Sun}} (\cos2\theta_n^-
        -\cos2\theta_0)
                    \\ +\frac{0.5\text{year}}{T_{\text{night}}} \cos2\theta_{\text{Sun}}\sin2\theta_0
                                    \sum\limits_{s=\pm1}
                                    w(\varsigma_s, \tau_{\text{midnight}}(\varsigma_s))
                                    \sum\limits_{j=1}^{n-1} \sDelta\theta_j
                                    \frac{\vartheta\big(r_j - r_n\sin(\chi+s\varepsilon)\big)}{2\pi\sqrt{\sin\varepsilon \cos\chi \sin(\chi+s\varepsilon)}}
                                    \frac{\sqrt{r_j^2/r_n^2 - \sin^2(\chi+s\varepsilon)}}{\lambda
                                    L_{n,j}(\chi + s \varepsilon)}
                    \\
                                    \times \cos\bigl\{2\sDelta\psi_{n,j}(\chi+s\varepsilon; E) + s'(s-1) \frac{\pi}{4}\bigr\}.
    \end{multline}
    Here, $\varsigma_{\pm1} = \pi,0$ correspond to the summer and the winter solstices, respectively; the daytime neutrino observation
    probability does not depend on the nadir angle (see Eq.~\eqref{valleyCliff}) and is thus unaffected by weighting,
    \begin{equation}
        \langle P_{\nu_e}(\text{day}; E) \rangle_w = P_{\nu_e}(\text{day}; E) \equiv \frac12 (1 +  \cos2\theta_{\text{Sun}}\cos2\theta_0).
    \end{equation}

    \vspace{0.5em}
    Let us now discuss the seasonal localization \eqref{locScale_sj} of the effect near the solstices.
    First of all, we emphasize that expression \eqref{locScale_sj}
    is based on the quadratic approximation for the phase $2\sDelta\psi_{n,j}(\Theta_{\text{N}}^{\text{midnight}}(\varsigma); E)$
    near the stationary point $\varsigma = \varsigma_s$ (cf. the general formula \eqref{locScale_def}
    with its quadratic approximation \eqref{locScale_quadratic}).
    This approximation is accurate when the second derivative
    $2\pd_\varsigma^2 \sDelta\psi_{n,j}(\Theta_{\text{N}}^{\text{midnight}}(\varsigma); E)$ is not too small at $\varsigma = \varsigma_s$
    (see Sec.~\ref{sec:Localization}). However, as one can learn, e.g., from \eqref{locScale_sj}, in the northern hemisphere,
    $\pd_\varsigma^2 \sDelta\psi_{n,j}(\Theta_{\text{N}}^{\text{midnight}}(\varsigma); E) \to 0$ for the winter solstice at the tropic ($\varsigma = 0$, $\chi \to \varepsilon+0$).
    This implies that both the estimation for the localization scale \eqref{locScale_sj} of the winter solstice stationary point
    and the asymptotic value of the integral \eqref{deltaP_year_w} are accurate not too close to the tropic.

    We have included selected values of $\delta{t}_{s,j}^{\text{solstice}}$ for the winter and summer solstice stationary points
    in Table~\ref{tab:locScale}, for a tropical latitude of the neutrino detector $\chi = 26\degree$, as well
    as for the Borexino ($\chi = 42.5\degree$) and the Super-Kamiokande ($\chi = 36.2\degree$) detectors.
    The values presented in Table~\ref{tab:locScale} correspond to the interfaces with the most definitive density
    jumps inside the Earth, i.e., with the largest mixing angle jumps $\sDelta\theta_j$. As a result,
    these interfaces are able to generate significant contributions
    to the sum entering the expression for the nighttime neutrino observation probability, see Eq.~\eqref{deltaP_year_w}.
    Note also that all interfaces but the Earth's surface ($r_j=r_n$) enter the sum in Eq.~\eqref{deltaP_year_w}
    twice, with $s' = +1$ and $s' = -1$. While we have included in Table~\ref{tab:locScale} the localization of both terms,
    the more significant of the two is the one with the lesser denominator $L_{n,j}$ in Eq.~\eqref{deltaP_year_w}.
    This denominator suppresses contributions of `distant' interfaces.

    As earlier, we have also included in Table~\ref{tab:locScale} the exposure time analogous to \eqref{delta_t_j}
    but measuring the duration of the season in which the Sun descends to shine through the $j$th interface at night,
    \begin{equation}\label{delta_t_j_season}
        \sDelta{t}^{\text{season}}_j \equiv \frac{1~\text{year}}{2\pi}  \int\limits_0^{2\pi}
                                \vartheta\left(\arcsin\frac{r_j}{r_n} - \Theta_{\text{N}}^{\text{midnight}}(\varsigma)\right)\;\diff\varsigma
                                = \frac{1~\text{year}}{2\pi}
                                   \times 2\mathcal{F}\left(
                                                \frac{\sin(\chi - \arcsin(r_j/r_n))}
                                                     {\sin\varepsilon}
                                   \right),
    \end{equation}
    where the function $\mathcal{F}$ is defined in \eqref{arccos_modified}. Quite analogously to the localization near midnights,
    the localization effect of the two solstices is characterized by the ratio $\sDelta{t}^{\text{season}}_j / \delta{t}_{s,j}^{\text{solstice}}$.

    In particular, one can see from Table~\ref{tab:locScale} that if one places the detector near the Tropic ($\chi = 26\degree$),
    the time localization scale corresponding to the contribution from the densest inner core is indeed smaller
    than the exposure time of this layer, so that one can isolate the contribution by reducing the season of observation,
    contracting it to about two months around
    the winter solstice (for ${}^8$B neutrinos). For ${}^7$Be neutrinos ($E = 862~\text{keV}$) having a 10-times smaller
    oscillation length, the localization time is $\sqrt{10} \sim 3$ times smaller, which, in principle, could also be
    useful for observing them, despite the challenging low-energy neutrino detection techniques and their fainter
    interaction with the Earth's matter. Indeed, high monochromaticity of ${}^7$Be neutrinos seems to favor possible
    detection of seasonal effects in them, when the sensitivities of detectors have grown enough to catch the corresponding
    matter effects \cite{Smirnov2015}. In what follows, however, we put our primary focus on the
    study of ${}^8$B neutrinos, pointing out that, in principle, the technique developed in the present paper
    can be directly applied to ${}^7$Be neutrinos.

    \begin{table}[tbh]
        \begin{tabular}{|c|c|c|c|c|c|}
            \hline
            Latitude & \multirow{2}{*}{Interfaces crossed} & $r_j$, & $\sDelta{t}^{\text{season}}_j$,
            & $\delta{t}^{\text{solstice}}_{-1,j}$ (winter solstice), month & $\delta{t}^{\text{solstice}}_{+1,j}$ (summer solstice), month \\
            $\chi$ &  & km & month & for $E = 10~\text{MeV}$ ($E = 862~\text{keV}$) & for $E = 10~\text{MeV}$ ($E = 862~\text{keV}$) \\
            \hline
            \multirow{7}{*}{$26.0\degree$} & atmosphere$\to$crust & 6371 & 12.0 & $3.1$ ($0.9$) & $0.8$ ($0.2$)  \\
            \cline{2-6}
            & upper mantle$\to$lower mantle & 5701 & 12.0 & $3.0$ ($0.9$) & $0.7$ ($0.2$)  \\
            \cline{2-6}
            & lower mantle$\to$outer core & 3480 & 7.2 & $2.6$ ($0.8$) & N/A \\
            \cline{2-6}
            & outer core$\to$inner core & 1221 & 3.3 & $1.7$ ($0.5$) & N/A \\
            \cline{2-6}
            & inner core$\to$outer core & 1221 & 3.3 & $2.1$ ($0.6$) & N/A \\
            \cline{2-6}
            & outer core$\to${}lower mantle & 3480 & 7.2 & $4.7$ ($1.4$) & N/A   \\
            \cline{2-6}
            & lower mantle$\to$upper mantle & 5701 & 12.0 & $13$ ($3.7$) & $1.7$ ($0.5$)   \\
            \hline
            \multirow{5}{*}{$36.2\degree$} & atmosphere$\to$crust & 6371 & 12.0 & $1.4$ ($0.4$) & $0.7$ ($0.2$) \\
            \cline{2-6}
            & upper mantle$\to$lower mantle & 5701 & 12.0 & $1.3$ ($0.4$) & $0.6$ ($0.17$)  \\
            \cline{2-6}
            & lower mantle$\to$outer core & 3480 & 5.5 & $1.1$ ($0.3$) & N/A\\
            \cline{2-6}
            & outer core$\to${}lower mantle & 3480 & 5.5 & $2.0$ ($0.6$) & N/A \\
            \cline{2-6}
            & lower mantle$\to$upper mantle & 5701 & 12.0 & $5.6$ ($1.6$) & $0.9$ ($0.3$)  \\
            \hline
            \multirow{5}{*}{$42.5\degree$} & atmosphere$\to$crust & 6371 & 12.0 & $1.1$ ($0.3$) & $0.7$ ($0.20$) \\
            \cline{2-6}
            & upper mantle$\to$lower mantle & 5701 & 10.3 & $1.1$ ($0.3$) & N/A   \\
            \cline{2-6}
            & lower mantle$\to$outer core & 3480 & 4.4 & $0.9$ ($0.3$) & N/A \\
            \cline{2-6}
            & outer core$\to${}lower mantle & 3480 & 4.4  & $1.5$ ($0.4$) & N/A   \\
            \cline{2-6}
            & lower mantle$\to$upper mantle & 5701 & 10.3 & $4.4$ ($1.3$) & N/A   \\
            \hline
        \end{tabular}
        \caption{Time localization scales $\delta{t}^{\text{solstice}}_{s,j}$ (see Eq.~\eqref{locScale_sj})
        and exposure times $\sDelta{t}^{\text{season}}_j$ (see Eq.~\eqref{delta_t_j_season})
        for various interfaces $j$ inside the Earth for neutrinos with $E = 10~\text{MeV}$ and
        $E = 862~\text{keV}$. ``N/A'' means that the Sun never descends low enough in summer so that
        neutrinos went through a given interface. The localization times are found in the quadratic approximation \eqref{locScale_quadratic}}
        \label{tab:locScale}
    \end{table}

    \vspace{0.5em}
    An interesting point we address now is what is happening at the tropic $\chi = \varepsilon$.
    The approximate expression \eqref{deltaP_year_w} for the weighted average
    predicts augmentation of the contribution of the winter stationary point near the tropic, due to  $\sqrt{\sin(\chi + s \varepsilon)}, s=-1$
    in the denominator. At the same time, somewhere close to the tropic, approximation \eqref{deltaP_year_w} clearly becomes inaccurate, since
    it diverges if one substitutes $\chi = \varepsilon$ in it.
    This divergence, however, is purely formal and is nothing but an indication of the fact that at the tropic,
    the winter solstice $\varsigma = 0$ becomes a degenerate stationary point such that
    \begin{eqnarray}
        \left.\pd_\varsigma^k \bigl( 2\sDelta\psi_{n,j}(\Theta_{\text{N}}^{\text{midnight}}(\varsigma); E) \bigr) \right|_{\varsigma = 0} &=& 0,
        \quad k = 1,2,3; \\
        \left.\pd_\varsigma^4 \bigl( 2\sDelta\psi_{n,j}(\Theta_{\text{N}}^{\text{midnight}}(\varsigma); E) \bigr) \right|_{\varsigma = 0}
        &=& -6\tan^2\varepsilon \times \lambda r_n (1 + s' r_n / r_j)\ne 0,
    \end{eqnarray}
    and it is illegal to use the quadratic approximation for the phase $2\sDelta\psi_{n,j}(\Theta_{\text{N}}^{\text{midnight}}(\varsigma); E)$
    around it. Using the forth-order approximation instead, we arrive at the localization scale for the winter solstice
    at the tropic $\chi = \varepsilon$ (cf.~\eqref{locScale_quartic})
    \begin{equation}\label{locScale_tropic}
        \delta{t}^{\text{solstice}}_{-1,j} = \frac{1~\text{year}}{2\pi}
                    \left[\frac{48\pi}{|\pd_\varsigma^4 \bigl( 2\sDelta\psi_{n,j}(\Theta_{\text{N}}^{\text{midnight}}(\varsigma); E)
                    \bigr)|}\right]^{1/4}_{\varsigma = 0}
                    = \left[\frac{8\pi}{\lambda r_n (r_n / r_j + s') \; \tan^2\varepsilon} \right]^{1/4}
                                \frac{1~\text{year}}{2\pi}, \quad \chi = \varepsilon.
    \end{equation}
    Note that even at the tropic, the summer solstice stationary point of the seasonal integral \eqref{sum365toIntegral} remains non-degenerate,
    $\pd_\varsigma^2 (2\sDelta\psi_{n,j}) \ne 0$, thus, one can use the expression \eqref{locScale_sj} for the corresponding localization scale
    $\delta{t}^{\text{solstice}}_{+1,j}$. The integral over the night \eqref{I_d} also contains only one non-degenerate stationary
    point at midnight, even when $\chi = \varepsilon$.

    Selected values of the above tropical localization are presented in Table~\ref{tab:locScale_Tropic}. It is worth pointing
    out that the time localization presented in it is substantially tighter than that calculated in the leading approximation for $\chi = 26\degree$
    and presented in Table~\ref{tab:locScale}, especially for the outer layers of the Earth. It is worth saying that
    the localization scales for the JUNO detector ($\chi = 22.5\degree$) are very close to those in Table~\ref{tab:locScale_Tropic}.
    \begin{table}[tbh]
        \begin{tabular}{|c|c|c|c|c|}
            \hline
            Latitude & \multirow{2}{*}{Interfaces crossed} & $r_j$ & $\sDelta{t}^{\text{season}}_j$,
            & $\delta{t}^{\text{solstice}}_{-1,j}$ (winter solstice), month\\
            $\chi$ & & km & month & for $E = 10~\text{MeV}$ ($E = 862~\text{keV}$)\\
            \hline
            \multirow{7}{*}{$\varepsilon = 23.4\degree$} & atmosphere$\to$crust & 6371 & 12.0 & $1.9$ ($1.1$)  \\
            \cline{2-5}
            & upper mantle$\to$lower mantle & 5701 & 12.0 & $1.9$ ($1.0$)   \\
            \cline{2-5}
            & lower mantle$\to$outer core & 3480 & 7.7 & $1.8$ ($1.0$) \\
            \cline{2-5}
            & outer core$\to$inner core & 1221 & 3.8 & $1.5$ ($0.8$) \\
            \cline{2-5}
            & inner core$\to$outer core & 1221 & 3.8 & $1.6$ ($0.9$) \\
            \cline{2-5}
            & outer core$\to${}lower mantle & 3480 & 7.7 & $2.4$ ($1.3$)   \\
            \cline{2-5}
            & lower mantle$\to$upper mantle & 5701 & 12.0 & $4.0$ ($2.1$)  \\
            \hline
        \end{tabular}
        \caption{Time localization scales $\delta{t}^{\text{solstice}}_{-1,j}$ (see Eq.~\eqref{locScale_tropic})
                and exposure times $\sDelta{t}^{\text{season}}_j$ (see Eq.~\eqref{delta_t_j_season})
                for a detector at the tropic $\chi = \varepsilon$ (see Eq.~\eqref{locScale_tropic})
                for various interfaces $j$ inside the Earth for neutrinos with $E = 10~\text{MeV}$ and $E = 862~\text{keV}$.
                The localization times are found in the fourth-order approximation \eqref{locScale_quartic},
                since the quadratic approximation presented in Table~\ref{tab:locScale} is not applicable to the tropic}
        \label{tab:locScale_Tropic}
    \end{table}

    \vspace{0.5em}
    From the above considerations, we conclude that the localization effect is indeed present in the day-night asymmetry
    of solar neutrinos, and opens a possibility to observe the contributions of inner Earth's layers (see Eq.~\eqref{deltaP_year_w})
    via concentration of the observations (i.e., the weighting function $w(\varsigma, \tau$)) around the midnights daily and around the two solstices seasonally.
    Moreover, the contribution of the winter solstice becomes even more distinguishable for detectors near the tropic.

    \vspace{0.5em}
    Since concentration of the observations is able to magnify the contributions of the stationary points, we would like to draw attention
    to the oscillatory behavior of these contributions to weighted time averages over the night \eqref{I_d_result}
    and over the year \eqref{deltaP_year_w}, as functions of the neutrino energy $E$. Such a behavior originates in the cosine functions
    containing the oscillation phases $2\sDelta\psi_{n,j}$ in the stationary points (the midnights and the solstices, respectively), which, in turn,
    depend on the oscillation length, i.e., on $E$. It is easy to observe that the positions of the peaks of the corrections
    to the average day-night asymmetry \eqref{deltaP_year_w} coming from the inner Earth's layers are approximately determined by
    the maxima (or minima) of the oscillating cosines. Namely, for the annual average, these peaks arise where
    \begin{equation}\label{E_N_eqn}
          E = E_{Nsj}, \qquad 2\sDelta\psi_{n,j}(\chi+s\varepsilon; E_{Nsj}) + s'(s-1) \frac{\pi}{4} = 2\pi N,
          \quad
          N \in \mathds{Z} \text{ or } N + \frac12 \in \mathds{Z},
    \end{equation}
    which leads to the distance between two adjacent maxima (minima)
    \begin{equation}\label{deltaE_N}
        \frac{\sDelta{E}}{E} \equiv \frac{E_{Nsj} - E_{N-1,sj}}{E_{Nsj}} \simeq \frac{\ell_{\text{osc}}}{L_{n,j}(\chi+s\varepsilon)} =
        \frac{\ell_{\text{osc}}}{r_n \cos(\chi + s \varepsilon) + s' \sqrt{r_j^2 - r_n^2 \sin^2(\chi + s \varepsilon)}}.
    \end{equation}
    In the implicit expression \eqref{E_N_eqn} for $E_N$, it is illegal to neglect the matter effect and set
    $2\sDelta\psi_{n,j} \approx 2\lambda L_{n,j}$, because for large $L_{n,j} \sim r_{\text{Earth}}$,
    such an approximation does not ensure the expected accuracy of $E_N$, which should be not worse than $\sDelta{E}$.
    A more or less accurate approximation based on the expansion of the matter term
    $\omega(x; E) \approx 1 - \eta(x; E) \cos 2 \theta_0$ (see Eq.~\eqref{omega_def}) leads to
    \begin{equation}\label{E_N_LO}
           E_{Nsj} \approx \frac{\sDelta{m}^2 \Bigl\{ r_n \cos(\chi + s \varepsilon) + s' \sqrt{r_j^2 - r_n^2 \sin^2(\chi + s \varepsilon)} \Bigr\}}
                     {4\pi N - s'(s-1)\pi/2 + 2\sqrt2 G_{\text{F}} \cos{2\theta_0} \int_{x_j}^{x_n} N_e(x) \diff{x}}.
    \end{equation}
    For 10-MeV solar neutrinos, the oscillation length is about 300~km, and for the contribution of the core-mantle interface ($r_j = 3480~\text{km}$) and $s' = -1$,
    the distance between two adjacent maxima $\sDelta{E} \sim 1~\text{MeV}$. Therefore, if a future detector becomes capable of resolving an
    oscillatory structure of the energy spectrum of the day-night asymmetry with periods of the order of 1~MeV,
    this will definitely provide a way to precisely determine the mass-squared difference $\sDelta{m^2}$. The more maxima the detector observes,
    the better the accuracy of the determination of $\sDelta{m}^2$. In principle, the interference condition \eqref{E_N_eqn} can also be used to determine
    the radius of the interface $r_j$.

    \vspace{0.5em}
    Finally, we would like to discuss the definition of the day-night asymmetry factor to be used for measuring it
    in the context of temporally weighted observations. The usual definition \cite{BaltzWeneser, BahcallKrastev, SuperK_DNA_latest}
    \begin{equation}\label{Adn_exp_year}
        A_{\text{dn}}^{\text{(year)}} = \frac{2(N_{\text{night}} - N_{\text{day}})}{N_{\text{night}} + N_{\text{day}}}
    \end{equation}
    has little notion in our case, since the total numbers of nighttime/daytime neutrino events $N_{\text{night,day}}$ are
    proportional to the total duration of night/day over the chosen observation term. The latter are equal over the whole year,
    while in summer, for instance, days last longer than nights. A natural generalization of \eqref{Adn_exp_year}
    applicable to temporally-weighted observations could be
    \begin{equation}\label{Adn_exp_w_0}
        A_{\text{dn}}^{(w)} = \frac{2(N_{\text{night}}^{(w)} / T_{\text{night}} - N_{\text{day}}^{(w)} / T_{\text{day}})}
                                    {N_{\text{night}}^{(w)} / T_{\text{night}}+ N_{\text{day}}^{(w)} / T_{\text{day}}},
    \end{equation}
    where $T_{\text{night,day}}$ are the weighted nighttime and daytime over the chosen season (see Eq.~\eqref{T_nightday}) and
    $N^{(w)}_{\text{night,day}}$ are weighted numbers of neutrino events observed during the night and the day, respectively.
    Still, the above definition of the day-night asymmetry factor is not optimal for experimental purposes
    because it includes the daytime and the nighttime terms on equal footing. Indeed, daytime events are no use weighting
    since we do not expect any localization effects involved. Moreover, one can use the data on the number of daytime
    events not only over the whole year but over $Y$ previous years of operation of the experiment as well,
    in order to find the average daytime event \emph{rate} $\dot{\bar{N}}_{\text{day}} = \frac{N_{\text{day}}}{(Y/2)~\text{years}}$
    and compare the latter with its \emph{weighted} nighttime counterpart $N_{\text{night}}^{(w)} / T_{\text{night}}$. Such an approach obviously reduces the
    statistical uncertainty of the day-night asymmetry factor. Therefore, within the present paper we will use the following
    definition of the asymmetry factor
    \begin{equation}\label{Adn_exp_w}
        A_{\text{dn}}^{(w)} = \frac{2(N_{\text{night}}^{(w)} / T_{\text{night}} - \dot{\bar{N}}_{\text{day}})}
                                    {N_{\text{night}}^{(w)} / T_{\text{night}}+ \dot{\bar{N}}_{\text{day}}}
    \end{equation}
    and assume that the daytime rate $\dot{\bar{N}}_{\text{day}}$ has been calculated over quite a long observation period
    and has a vanishing statistical uncertainty. Some types of the systematics may cancel in the numerators
    of \eqref{Adn_exp_w_0}, \eqref{Adn_exp_w}  as well. 
    Finally, it should be emphasized that event numbers $N_{\text{night}}^{(w)}$,
    $\dot{\bar{N}}_{\text{day}}$ represent both the oscillations of solar neutrinos and the experimental setup,
    and the simplest model of the latter could be in the manner of Eqs.~\eqref{Nnight}, \eqref{Nnight_variance}. Therefore,
    it is only in the case of a perfect energy-resolving neutrino experiment with narrow energy bins
    that the above definition of the day-night asymmetry factor $A_{\text{dn}}^{(w)}$
    is proportional to the `asymmetry spectrum' which we will use in our simulations below,
    \begin{equation}\label{Adn_exp_w_theor}
        \hat{A}_{\text{dn}}^{(w)}(E) = \frac{2\bigl[\avg{P(\text{night}; E)}_w - P(\text{day}; E)\bigr]}
                                            {\avg{P(\text{night}; E)}_w + P(\text{day}; E)}.
    \end{equation}
    The only and the principal advantage of the latter asymmetry factor is that it is defined independently of the neutrino
    detection mechanism. On the other hand, in order to conform our results in terms of $\hat{A}_{\text{dn}}^{(w)}(E)$
    to a certain detection and data processing scheme, one should make a transformation from the neutrino energy $E$
    to the electron recoil energy $T$ (or any other relevant observable experimental parameter) and also take into
    account the assumed energy spectrum of solar neutrinos and the detector sensitivity curve. Finally, one should
    distribute the recoil energy values among a number of bins, as one usually does.

    \vspace{0.5em}
    Expressing our estimation of the day-night asymmetry \eqref{deltaP_year_w} in terms of the factor \eqref{Adn_exp_w_theor}
    just introduced, we finally obtain
    \begin{multline}\label{Adn_analytical_w}
        \hat{A}_{\text{dn}}^{(w)}(E) \approx \frac{\cos2\theta_{\text{Sun}} (\cos2\theta_n^- - \cos2\theta_0)}{1 + \cos2\theta_0 \cos2\theta_{\text{Sun}}}
                    + \frac{\cos2\theta_{\text{Sun}}\sin2\theta_0}{1 + \cos2\theta_0 \cos2\theta_{\text{Sun}}}
                    \\ \times \frac{0.5\text{year}}{T_{\text{night}}}
                                    \sum\limits_{s=\pm1}
                                    w(\varsigma_s, \tau_{\text{midnight}}(\varsigma_s))
                                    \sum\limits_{j=1}^{n-1} \sDelta\theta_j
                                    \frac{\vartheta\big(r_j - r_n\sin(\chi+s\varepsilon)\big)}{2\pi\sqrt{\sin\varepsilon \cos\chi \sin(\chi+s\varepsilon)}}
                                    \frac{\sqrt{r_j^2/r_n^2 - \sin^2(\chi+s\varepsilon)}}{\lambda
                                    L_{n,j}(\chi + s \varepsilon)} \\
                                    \times \cos\bigl\{2\sDelta\psi_{n,j}(\chi+s\varepsilon; E) + s'(s-1) \frac{\pi}{4}\bigr\}.
    \end{multline}
    The approximations that have led to this expression are the adiabaticity, the piecewise continuous Earth's density profile,
    the smallness of the asymmetry (i.e. the effective angle jumps $\sDelta\theta_j, \ \theta_n^- - \theta_0$ should be considerably smaller than unity).
    Moreover, we have used the leading-order stationary phase approximation which is applicable for detector latitudes $\chi > \varepsilon$,
    however, not too close to the tropic, for the weighting functions $w(\varsigma,\tau)$ that are almost constant within the time localization scale,
    and for interfaces which lie much deeper than the oscillation length under the detector,
    $L_{n,j}(\chi + s \varepsilon) \gg \pi / \lambda$. Nevertheless, as it was discussed in \cite{DNA_Kharlanov_PRD}, non-trivial structure
    of the Earth's density profile within much less than one oscillation length does not affect regeneration of neutrinos. In particular,
    for ${}^8$B neutrinos, several density jumps within the Earth's crust under the detector
    (which are not deep enough for $L_{n,j} \gg \pi / \lambda$ to hold)
    can be treated as a single jump on the exit from the Earth, without much loss of accuracy. As a result, the sum in the above expression should
    not include the density jumps in the Earth's crust under the neutrino detector.

    \section{Numerical simulation of the localization effects}\label{sec:NumericalSimulation}

    Let us now make a numerical simulation of the effects studied in the previous sections, namely,
    look at how an observation (or experimental data processing) restricted to a short period around the winter solstice and/or midnights can revive,
    i.e., magnify the contributions to the oscillation probability \eqref{deltaP_year_w} representing the effect
    of the inner layers of the Earth.

\begin{figure}[tbh]
    \includegraphics[width=7cm]{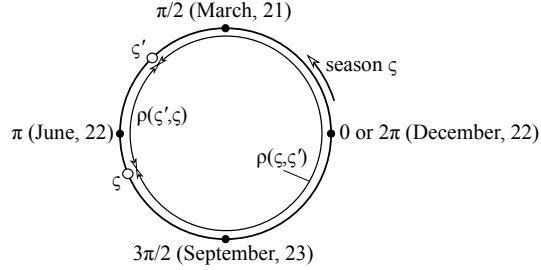}
    \caption{`Oriented distance' function $\rho(\varsigma,\varsigma')$ on a circle $\varsigma, \varsigma' \in [0,2\pi)$ used in the definitions
    \eqref{w_pseudoGaussian}, \eqref{w_pseudoGaussian_day} of the weighting functions}
    \label{fig:rho}
    \end{figure}

\begin{figure}[tbh]
    \includegraphics[width=10cm]{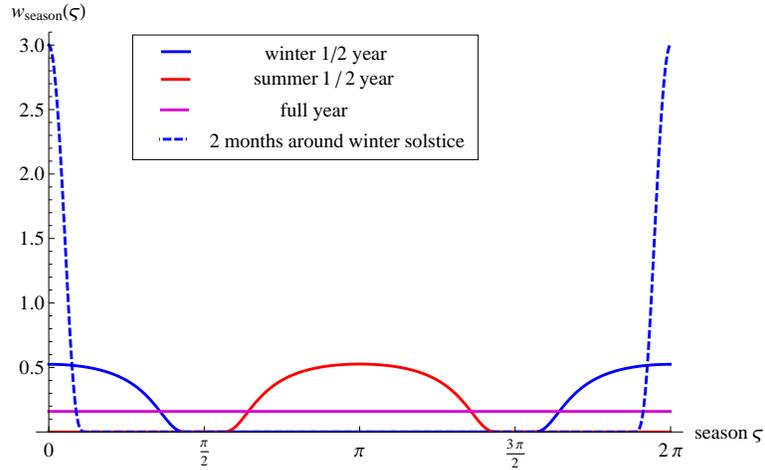}
    \caption{The full-year weighting function and weighting functions $w_{\text{season}}$ extracting different seasons
             (for the latter ones, edge width $\epsilon_\varsigma = \pi/6$, equivalent to 1~month)}
    \label{fig:wfs}
    \end{figure}

    In our simulations, we should bear in mind that the actual weighting function $w(\varsigma,\tau)$ to be used for
    evaluating the neutrino observation probability over a certain daytime/season should not have abrupt jumps,
    i.e., it should behave much like a bump function (an infinitely-differentiable function with a compact support).
    This is necessary to rely upon the stationary point approximation \eqref{statPhase} without the boundary term,
    as we did in the previous section. We employed the two following shapes of the weighting function
    \begin{eqnarray}\label{w_pseudoGaussian}
        w_{\text{season}}(\varsigma,\tau) &=&  \const \times \Omega(\varsigma; \; \varsigma_1, \varsigma_2, \epsilon_\varsigma),     \\
        \label{w_pseudoGaussian_day}
        w_{\text{season,day}}(\varsigma,\tau) &=&  \const \times \Omega(\varsigma; \; \varsigma_1, \varsigma_2, \epsilon_\varsigma)
                                               \Omega(\tau - \tau_{\text{midnight}}(\varsigma) \, \mathrm{mod} \, 2\pi; \; \tau_1, \tau_2, \epsilon_\tau),     \\
        \Omega(\xi; \xi _1, \xi _2, \epsilon) &\equiv&
                        \begin{cases}
                            0,                                                 &
                            (\xi_1 < \xi_2 \text { and } \xi \not\in [\xi_1, \xi_2])
                            \text{ or } ( \xi_1 > \xi_2 \text { and } \xi \in [\xi_2, \xi_1]), \\
                             \exp\left[ -\frac{\epsilon^2}{\rho^2(\xi, \xi_1)}
                                            -\frac{\epsilon^2}{\rho^2(\xi_2, \xi) }
                                    \right] & \text{otherwise},
                       \end{cases}
    \end{eqnarray}
    where $\varsigma, \tau \in [0,2\pi)$, $\varsigma_{1,2}, \tau_{1,2} \in [0,2\pi)$ are the start/end points
    of the observation season/hours, and the `oriented distance' function on a circle
    \begin{equation}\label{rho_def}
        \rho(\xi, \xi') = \begin{cases}
                                          \xi' - \xi, & 0 \le \xi \le \xi' < 2\pi, \\
                                          \xi' - \xi + 2\pi, & 0 \le \xi' < \xi < 2\pi
                                      \end{cases}
                                    = 2\pi - \rho(\xi', \xi)
    \end{equation}
    (see Fig.~\ref{fig:rho}). The function $w_{\text{season}}$ performs a seasonal averaging, while $w_{\text{season,day}}$ selects certain hours during
    every night as well. The definitions of $w_{\text{season}}$ and $w_{\text{season,day}}$ ensure their smoothness on a circle representing the year
    (i.e., on a segment $\varsigma \in [0, 2\pi)$ with two end points identified),
    rather than just on a segment. The edge width parameters $\epsilon_\varsigma$, $\epsilon_\tau$ are
    to be chosen empirically. Examples of the weighting function $w_{\text{season}}$ introduced are presented in Fig.~\ref{fig:wfs}.

    As usual, we calculated time averages of the form~\eqref{wAvg_def}
    using the solar exposure function $\epsilon(\Theta)$ at a given latitude $\chi$ and its weighted version
    $\epsilon_w(\Theta)$ \cite{BahcallKrastev, Astronomy, Lisi_seasonalVariations},
    \begin{gather}\label{expFunc}
    \epsilon(\Theta) = \int\limits_{1~\text{year}} \frac{\diff{t}}{1~\text{year}} \delta(\Theta_{\text{N}}(t) - \Theta), \qquad
    \epsilon_w(\Theta) = \int\limits_{1~\text{year}} \frac{w(t)\diff{t}}{1~\text{year}} \delta(\Theta_{\text{N}}(t) - \Theta); \\
    \int\limits_{0}^{\pi} \epsilon(\Theta)\diff\Theta = \int\limits_{0}^{\pi} \epsilon_w(\Theta)\diff\Theta = 1 \qquad
    \text{(the latter by virtue of Eq.~\eqref{w_normalization})},  \label{expFunc_norm}\\
    \langle P_{\nu_e}(\text{night};E) \rangle_{w} \equiv \int\limits_{\text{1~year}}\frac{w(t)\diff{t}}{T_{\text{night}}}\;
                                                                    \vartheta(\pi/2 - \Theta_{\text{N}}(t)) \; P_{\nu_e}(\Theta_{\text{N}}(t); E)
            = \frac{\int_0^{\pi/2} P_{\nu_e}(\Theta;E)\epsilon_w(\Theta)\diff\Theta}{\int_0^{\pi/2}\epsilon_w(\Theta)\diff\Theta}.
    \end{gather}
    The nighttime observation probabilities $P_{\nu_e}(\Theta_{\text{N}}, E)$ for various $\Theta_{\text{N}}$ have been computed via direct integration of
    the MSW equation \eqref{MSW_equation} inside the Earth using the PREM density profile \cite{PREM}. However,
    the result of direct integration agrees with the adiabatic approximation \eqref{valleyCliff} to a high accuracy,
    so the latter could have been used as well.

\begin{figure}[tbh]
    \includegraphics[width=12cm]{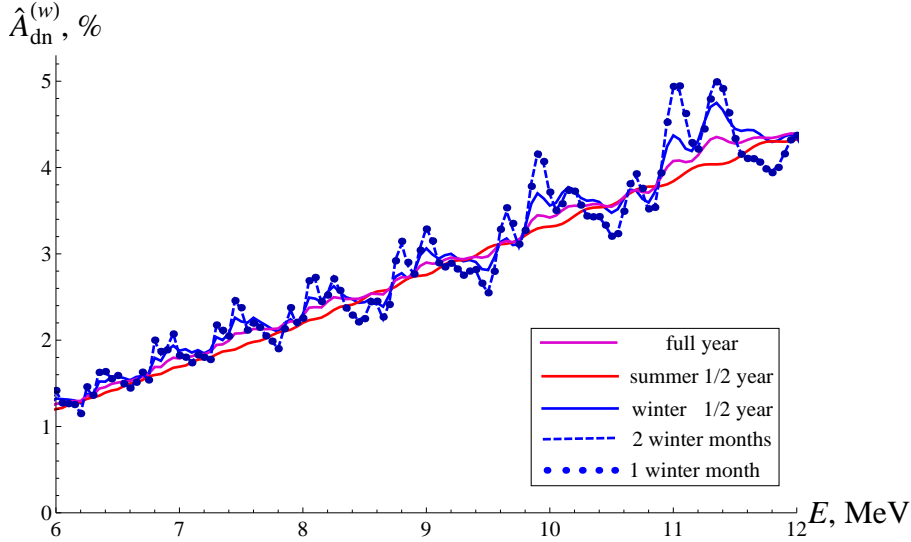}
    \caption{Day-night asymmetry factor $\hat{A}_{\text{dn}}^{(w)}(E)$ as a function of the neutrino energy $E$ for the detector latitude
        $\chi = 26\degree$ and different averaging seasons
        }
    \label{fig:Adn_E}
    \end{figure}

    \vspace{0.5em}
    Let us first have a look at the energy spectra of the day-night asymmetry factor $\hat{A}_{\text{dn}}^{(w)}(E)$
    for various observation terms. Fig.~\ref{fig:Adn_E} demonstrates these spectra for a year-long observation ($w(t) \equiv 1$),
    for two half-years centered in the winter and the summer solstices (edge width $\epsilon_\varsigma = 20\text{ days}$),
    for a 2-month term around the winter solstice (edge width $\epsilon_\varsigma = 10\text{ days}$), and for a 1-month term
    around the winter solstice (edge width $\epsilon_\varsigma = 5\text{ days}$). We have taken the latitude $\chi = 26\degree$ quite
    close to the tropic but not exactly at it, because our analytical considerations are not accurate too close to the tropic.
    From Fig.~\ref{fig:Adn_E}, it obviously follows that there is no use reducing the observation window to less than two months
    around the `Christmas' (winter solstice), since the results will be the same, while the number of events observed will be only smaller.
    This fact agrees with our estimation of the localization scale presented in Table~\ref{tab:locScale}. Further, the `smoothest'
    curve in Fig.~\ref{fig:Adn_E} (i.e. most free of oscillations) is the red one for the summer half-year. Indeed,
    the corresponding weighted day-night asymmetry does not contain the contribution of the winter solstice stationary point
    that is amplified near the tropic. As a result, the violet curve corresponding to the full-year (non-weighted) observation
    owes its (comparatively small) oscillations mainly to the winter part of the year. These oscillations are definitely amplified in
    the 2-month day-night asymmetry.

    \begin{figure}[tbh]
    $$
    \begin{array}{cc}
        \includegraphics[width=8cm]{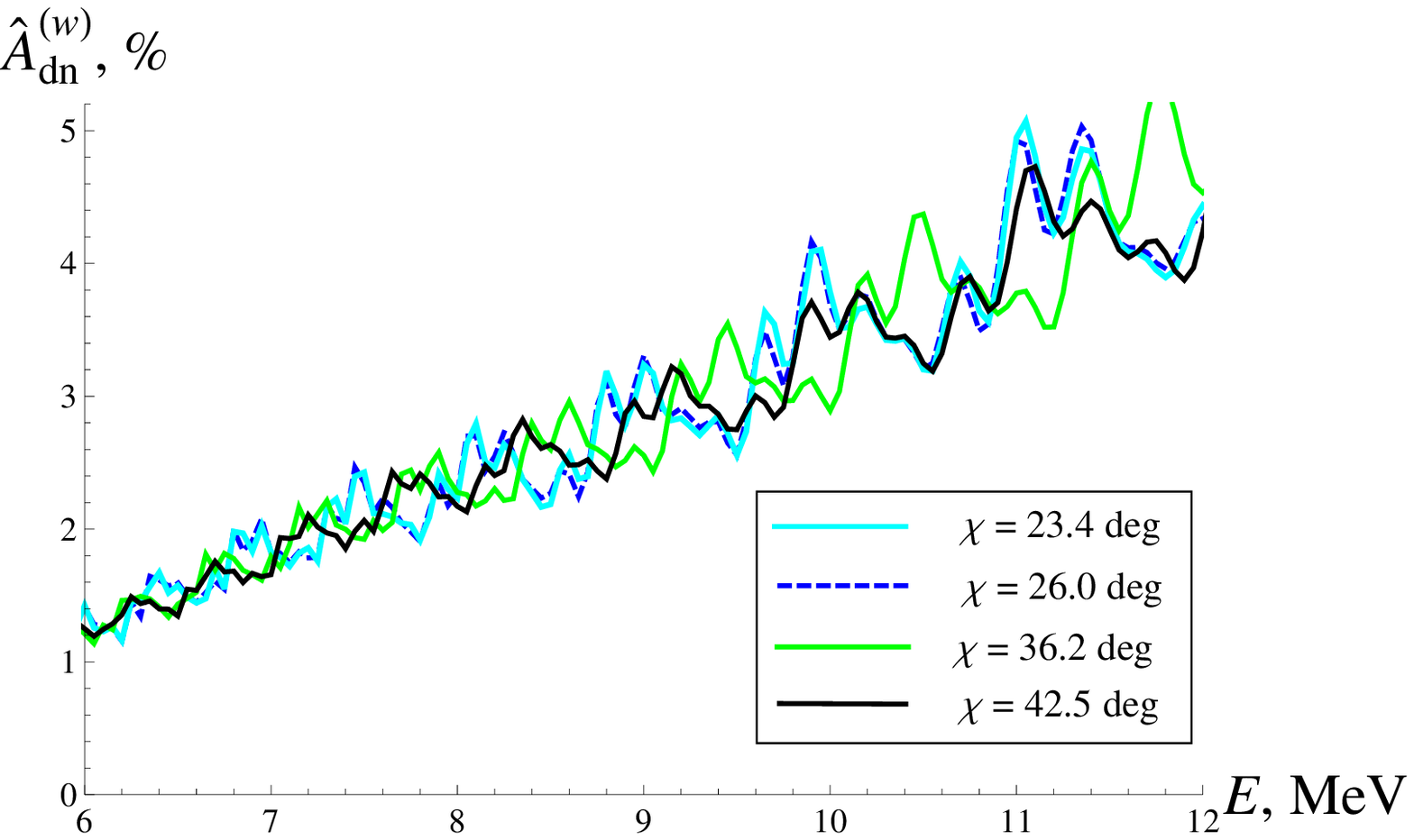}  &  \includegraphics[width=8.5cm]{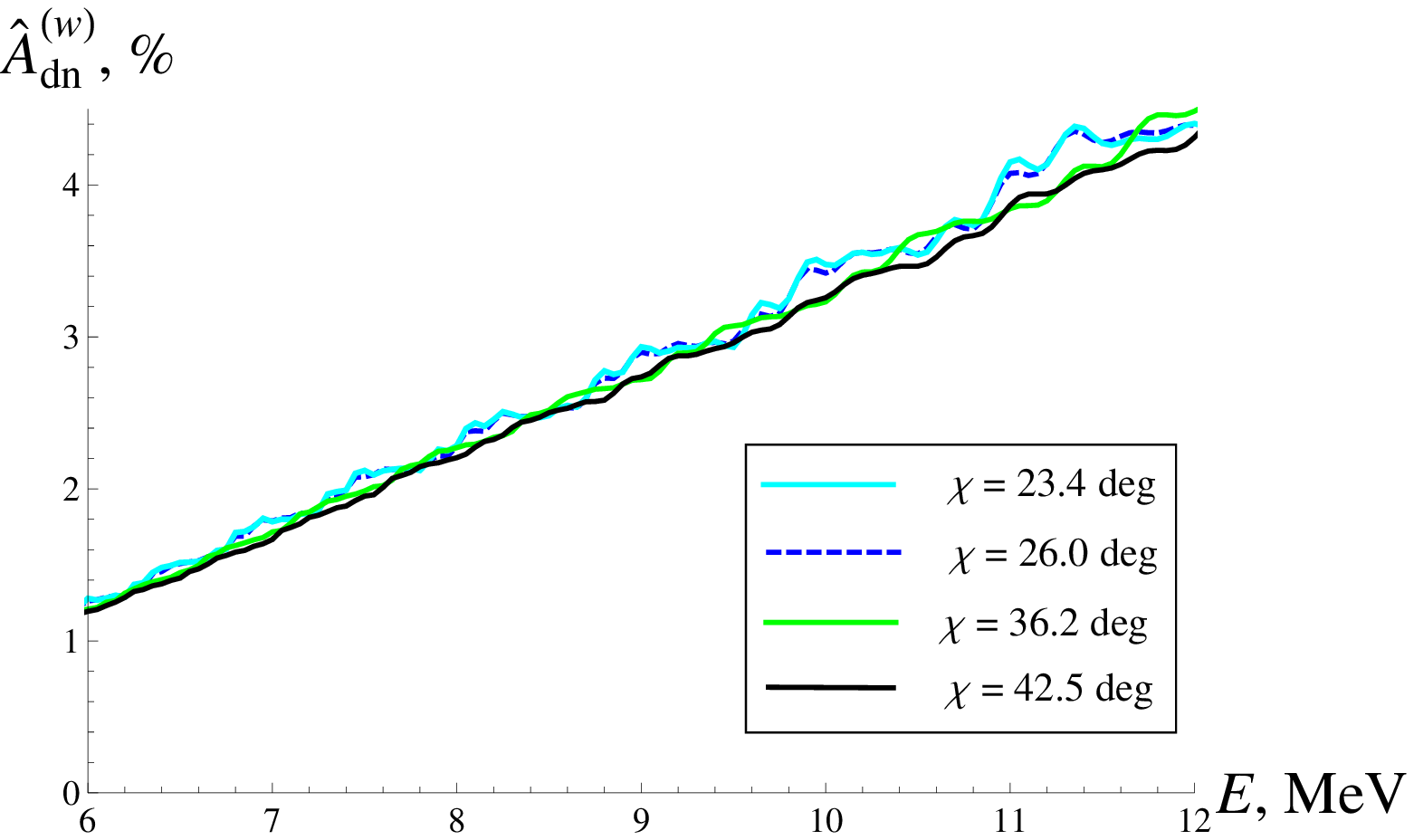} \\
        \text{(a)}                                         &  \text{(b)}
    \end{array}
    $$
    \caption{Day-night asymmetry factor $\hat{A}_{\text{dn}}^{(w)}(E)$ for different latitudes: (a) for the 2-month term around the winter solstice,
    (b) for the whole year (non-weighted)}
    \label{fig:Adn_E_latitudes}
    \end{figure}

    Next, from Fig.~\ref{fig:Adn_E_latitudes}a showing the two-month weighted asymmetries for latitudes $\chi = 23.4\degree, 26\degree, 36.2\degree, 42.5\degree$,
    we see that the effect at the tropic extends quite well to within at least 10 degrees north of it, while at the Borexino latitude
    $\chi = 42.5\degree$, it is considerably smaller. Note that the non-trivial signatures of the stationary points
    in the full-year effects (Fig.~\ref{fig:Adn_E_latitudes}b) decay much faster with the distance from the tropic.

\begin{figure}[tbh]
    \includegraphics[width=14cm]{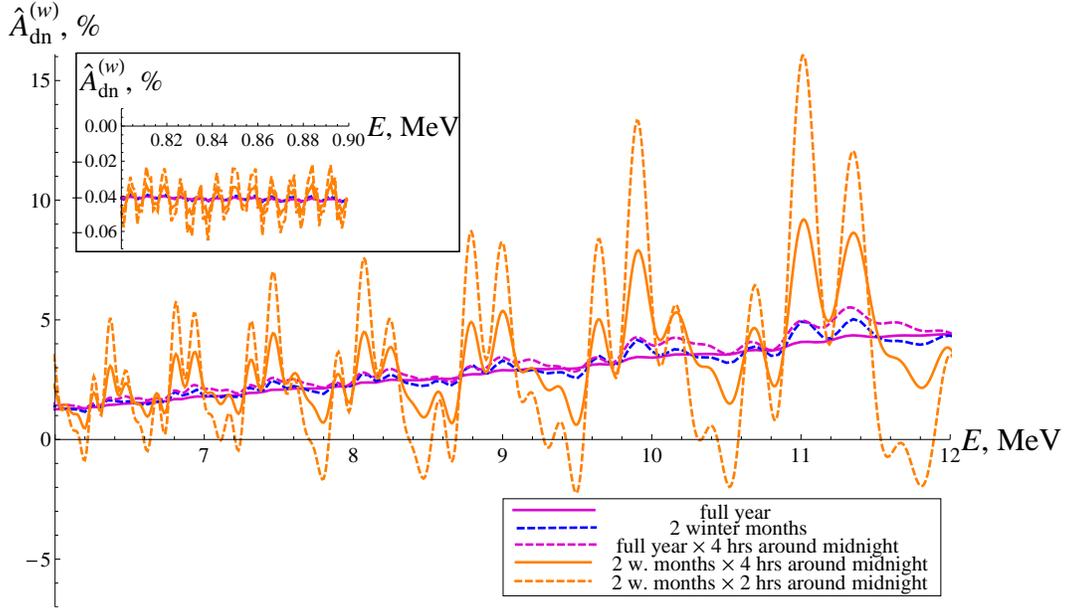}
    \caption{Day-night asymmetry factor $\hat{A}_{\text{dn}}^{(w)}(E)$ as a function of the neutrino energy $E$ for the detector latitude
        $\chi = 26\degree$ and different averaging seasons/hours: Solid violet curve for the whole year, dashed blue
        for 2 months around the winter solstice, dashed violet for 4 hours around midnights during the whole year,
        solid (dashed) orange curves for 4 (2) hours around midnights during 2 months around the winter solstice.
        Inset: the same curves in the ${}^7$Be energy band}
    \label{fig:Adn_E_midnightEffect}
    \end{figure}

    \vspace{0.5em}
    So far, we have investigated the properties of the seasonal localization using the corresponding weighting
    function $w_{\text{season}}(\varsigma, \tau)$. If we localize the observations around midnights as well, i.e.,
    use the weighting function $w_{\text{season,day}}(\varsigma, \tau)$ instead,
    then the oscillations on the $\hat{A}_{\text{dn}}^{(w)}(E)$ curves will get drastically amplified. Indeed,
    in Fig.~\ref{fig:Adn_E_midnightEffect}, one may observe that even a reduction of the nighttime observations made during
    the whole year to 4~hours around midnights results in an amplification comparable to that achieved by using
    a 2-month observation term around the winter solstice stationary point. A combination of the two types
    of weighting is even able to make the contributions of the stationary points dominate
    the cumulative contribution of the layer under the detector.

    In view of such a strong enhancement of the stationary point contributions, one may naturally propose that,
    if ${}^7$Be solar neutrinos ($E = 0.862~\text{MeV}$) happen to hit a peak in Fig.~\ref{fig:Adn_E_midnightEffect},
    then the observation of the day-night effect for such neutrinos (that has not been performed experimentally yet \cite{Borexino})
    could be strongly assisted by a time-localized data processing. For that reason, we have included in
    Fig.~\ref{fig:Adn_E_midnightEffect} an inset with the fragments of the curves in the low-energy domain in question.
    One observes that, even though an enhancement of the asymmetry may occur and, in principle, could provide a way
    of precision determination of $\sDelta{m^2}$, time-localized observation of ${}^7$Be neutrinos is unable
    to magnify the day-night effect strongly enough to considerably improve the (statistical) signal-to-noise ratio.
    Not to mention systematic uncertainties that constitute major obstacle on the way to observing this effect for ${}^7$Be neutrinos
    (see, e.g., \cite{Borexino}).

    \begin{figure}[tbh]
    \includegraphics[width=9cm]{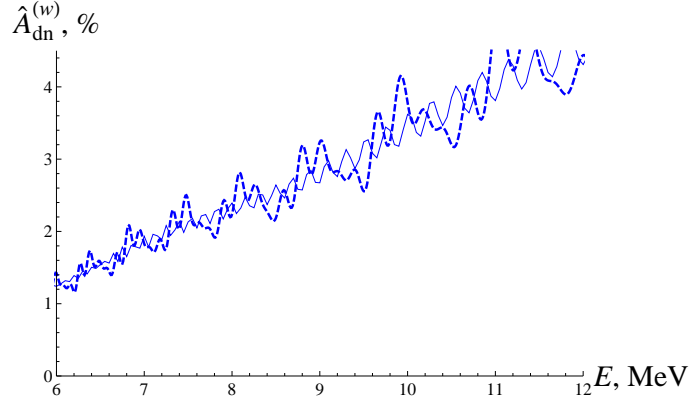}
    \caption{Day-night asymmetry factor $\hat{A}_{\text{dn}}^{(w)}(E)$ for two months around the winter solstice at
    a tropical detector ($\chi = 23.4\degree$) and two different density distributions inside the Earth,
    dashed curve for the PREM model, thin solid curve for the `homogeneous-core' distribution \eqref{PREM_noCore}}
    \label{fig:Adn_E_coreEffect}
    \end{figure}
    \begin{figure}[tbh]
    \includegraphics[width=9cm]{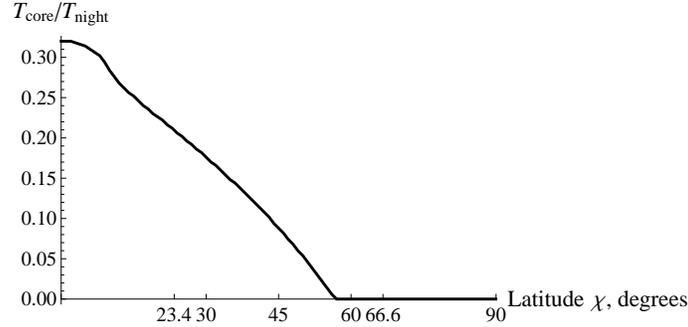}
    \caption{The part of nighttime $T_{\text{core}}/T_{\text{night}}$ (accumulated over the whole year)
    the detector will observe solar neutrinos passing through the Earth's core,
    depending on the latitude of the detector $\chi$}
    \label{fig:Tcore_chi}
    \end{figure}

    It is also worth demonstrating explicitly the contribution coming from the Earth's core comparing two simulations
    for the asymmetry at the tropic ($\chi = 23.4\degree$) with different Earth's density profiles. Namely,
    we have taken a PREM model modification which replaces the core by a homogeneous ball, eliminating
    the core-mantle discontinuity of the density,
    \begin{equation}\label{PREM_noCore}
        \tilde{N}_e(r) = \begin{cases}
                            N_e^{(\text{PREM})}(r), & r \ge r_{\text{core}}, \\
                            N_e^{(\text{PREM})}(r_{\text{core}} + 0), & r < r_{\text{core}},
                        \end{cases}
    \end{equation}
    $r_{\text{core}} \approx 3480~\text{km}$ and have plotted the asymmetry against the energy for such a density distribution.
    The result, together with the one within PREM, is shown in Fig.~\ref{fig:Adn_E_coreEffect} and clearly says that the dominant
    vertical oscillations of the curves in Figs.~\ref{fig:Adn_E},~\ref{fig:Adn_E_latitudes},~\ref{fig:Adn_E_midnightEffect} with the period about $1~\text{MeV}$
    originate in the Earth's core (in fact, it turns out that the remaining `no-core' oscillations in Fig.~\ref{fig:Adn_E_coreEffect}
    come from the entry point on the other side of the Earth). In this context, another plot also seems informative, namely,
    the total time during the year that the Sun shines through the Earth's core,
    \begin{equation}\label{T_core}
        T_{\text{core}} = \int\limits_{1~\text{year}} \diff{t} \, \vartheta\bigl(\arcsin(r_{\text{core}} / r_{\text{Earth}}) - \Theta_{\text{N}}(t)\bigr).
    \end{equation}
    We have depicted the dependence of $T_{\text{core}}$
    on the latitude in Fig.~\ref{fig:Tcore_chi}. One observes that, even though
    $T_{\text{core}}(\chi = 23.4\degree) \, / \, T_{\text{core}}(\chi = 36.2\degree) \approx 1.5$,
    the amplitudes of the oscillations of the curves $\hat{A}_{\text{dn}}^{(w)}(E)$ in Fig.~\ref{fig:Adn_E_latitudes}a
    corresponding to these two latitudes are almost equal. This is, in fact, a manifestation of the time localization
    of the effect for both latitudes (see Tables~\ref{tab:locScale},~\ref{tab:locScale_Tropic}).

    \vspace{0.5em}

\begin{figure}[tbh]
        \includegraphics[width=12cm]{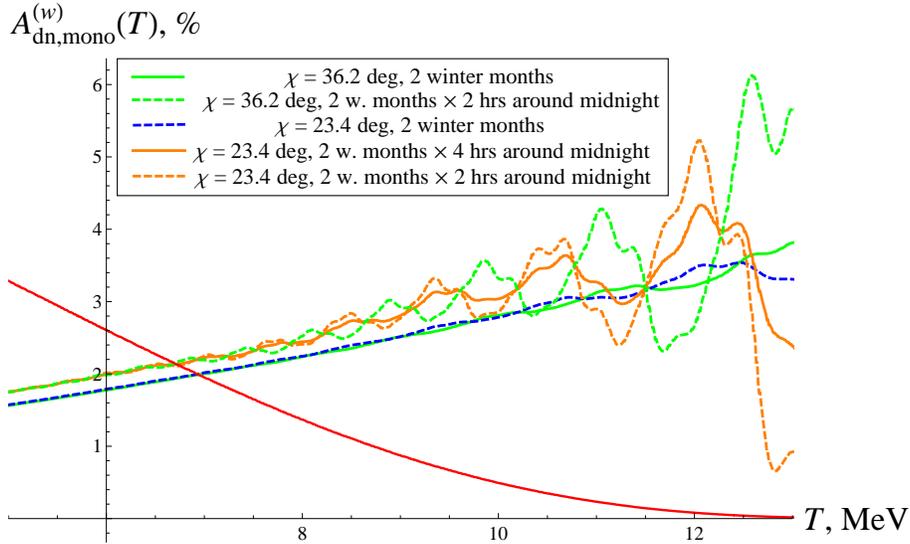}
        \caption{Day-night asymmetry factors $A_{\text{dn,mono}}^{(w)}(T)$ for narrow recoil energy bins for a detector at the latitude $\chi = 36.2\degree$
        (solid green curve) and a tropical detector ($\chi = 23.4\degree$, dashed blue curve and two orange curves);
        the observation terms are indicated in the legend. The solid red curve shows the number
        of electron scattering events per energy bin in arbitrary units (see the denominator in Eq.~\eqref{Adn_exp_w_simulation_mono})}
        \label{fig:Adn_T}
    \end{figure}

    Finally, we must admit that the neutrino energy is not a directly measurable quantity
    within a typical scattering detection scheme used in, e.g., water Cherenkov detectors.
    Thus, we have performed a recalculation of the day-night asymmetry in terms of the recoil electron energy. Namely, for an estimation, we took
    the tree-level Weinberg--Salam cross sections for the elastic $\nu_e e$ and $\nu_x e$ scattering processes \cite{Nu_e_CrossSection,Bahcall_CrossSections}
    and adopted a simple detection scheme with the sensitivities entering Eq.~\eqref{Nnight} taken in the form
    \begin{equation}\label{detection_sensitivity}
        \sigma_{\nu_{e,x}}(E) = \int\limits_{T_1}^{T_2} \frac{\diff\sigma_{\nu_{e,x}}(T;E)}{\diff{T}}\, \diff{T},
    \end{equation}
    where $[T_1, T_2]$ is the recoil energy bin and $\diff\sigma_{\nu_{e,x}}(T;E) / \diff{T}$ are the differential scattering cross sections.
    Then, assuming that the day-night asymmetry is much smaller than unity (which is the case for solar neutrinos),
    in the leading order we obtain, according to definition~\eqref{Adn_exp_w},
    \begin{equation}\label{Adn_exp_w_simulation}
        A_{\text{dn}}^{(w)} \approx \frac{N_{\text{night}}^{(w)}}{T_{\text{night}}\dot{\bar{N}}_{\text{day}}} - 1
                            \approx \frac{\int \Phi(E)\diff{E} \int_{T_1}^{T_2} \diff{T} \sDelta\frac{\diff\sigma(E,T)}{\diff{T}} P_{\text{day}}(E) \hat{A}_{\text{dn}}^{(w)}(E)}%
                                        {\int \Phi(E)\diff{E} \int_{T_1}^{T_2} \diff{T} \left\{
                                                                                    \sDelta\frac{\diff\sigma(E,T)}{\diff{T}} P_{\text{day}}(E) +
                                                                                    \frac{\diff\sigma_{\nu_x}(E,T)}{\diff{T}}
                                                                                \right\}
                                        },
    \end{equation}
    where $\sDelta\frac{\diff\sigma(E,T)}{\diff{T}} \equiv \frac{\diff\sigma_{\nu_e}(E,T)}{\diff{T}} - \frac{\diff\sigma_{\nu_x}(E,T)}{\diff{T}}$
    is the difference between cross sections for two neutrino flavors. Note that the denominator does not contain terms that could
    manifest the non-trivial effect of the solstices, since it does not contain the nighttime oscillation probabilities. Moreover,
    it is of use to formally consider a narrow bin $[T, T + \diff{T}]$ instead of $[T_1, T_2]$, arriving at the `monochromatic version'
    of the above asymmetry factor,
    \begin{equation}\label{Adn_exp_w_simulation_mono}
        A_{\text{dn,mono}}^{(w)}(T) \approx \frac{\int \Phi(E)\diff{E} \; \sDelta\frac{\diff\sigma(E,T)}{\diff{T}} P_{\text{day}}(E) \hat{A}_{\text{dn}}^{(w)}(E)}%
                                        {\int \Phi(E)\diff{E} \left\{
                                                                                    \sDelta\frac{\diff\sigma(E,T)}{\diff{T}} P_{\text{day}}(E) +
                                                                                    \frac{\diff\sigma_{\nu_x}(E,T)}{\diff{T}}
                                                                                \right\}
                                        }.
    \end{equation}
    The above, monochromatic asymmetry factors for a water Cherenkov detector at the latitude $\chi = 36.2\degree$ of Super~Kamiokande and
    for a detector at the tropic $\chi = 23.4\degree$ are presented in Fig.~\ref{fig:Adn_T}. In the calculations, we used the solar neutrino spectrum $\Phi(E)$
    predicted by the standard solar model \cite{Boron8Spectrum}. We also included in the graph the denominator of Eq.~\eqref{Adn_exp_w_simulation_mono},
    in order to estimate the expected recoil electron energy distribution in the detector.
    One observes that for the high-energy portion of solar neutrinos, corresponding to $T = 9-11~\text{MeV}$,
    the peak-to-peak amplitude of oscillations of the asymmetry factor almost reaches 1\%, which is comparable with the leading-order contribution.
    Moreover, if a detector is able to at least partially account for the angular distribution of recoil electrons,
    then the observed effect will become even more pronounced. Note also that the positions of the peaks on the $A_{\text{dn,mono}}^{(w)}(T)$ curve
    for $\chi = 36.2\degree$ are shifted relative to the one for a tropical detector, which is a manifestation of the interference condition \eqref{E_N_eqn}.
    Thus, in principle, these peaks can also be used as a tool for measuring the oscillation length of neutrinos,
    i.e., in other words, the $\sDelta{m^2}$ parameter. Moreover, since the oscillatory signatures in Fig.~\ref{fig:Adn_T}
    are virtually counter-phase for the two latitudes, one could combine the data collected at the two corresponding
    detectors to observe the difference effect two times larger than the individual effects at each of the detectors.
    This may also cancel certain systematics.

    \vspace{0.5em}
    In contrast to Fig.~\ref{fig:Adn_T}, for a facility with a perfect neutrino energy resolution
    located near the tropic, the energy dependence in Fig.~\ref{fig:Adn_E_midnightEffect} exactly corresponds to what is measured in the experiment,
    up to a factor of the form $1 + \frac{\sigma_{\nu_x}(E)}{P_{\text{day}}(E)\, (\sigma_{\nu_e}(E) - \sigma_{\nu_x}(E))}$ of the order of unity.
    Then, in order to observe the most vivid oscillations on the $\text{2-month}~\times~\text{2-hour}$ curve for $\hat{A}^{(w)}_{\text{dn}}(E)$,
    one could introduce two neutrino energy `bins', namely, the sets of $E$ where $\hat{A}^{(w)}_{\text{dn}}(E)$ should lie above
    the anomaly-free day-night asymmetry $\hat{A}^{(\text{regular})}_{\text{dn}}(E)$ (e.g., the summer curve)
    and where it should lie below the `regular' curve, and then collect the events in these two bins.
    A natural measure of the presence of oscillations as those seen in Fig.~\ref{fig:Adn_E_midnightEffect} would be
    \begin{equation}
    \Omega \equiv \sum\limits_{s=\pm1}s\;
                    \frac{\int \Phi(E)\diff{E} \; \vartheta\bigl[s(\hat{A}_{\text{dn}}^{(w)}(E) - \hat{A}^{(\text{regular})}_{\text{dn}}(E))\bigr]\;
                    \int_{T_{\text{min}}}^\infty \diff{T} \sDelta\frac{\diff\sigma(E,T)}{\diff{T}} P_{\text{day}}(E) \hat{A}_{\text{dn}}^{(w)}(E)
                        }%
                        {\int \Phi(E)\diff{E} \; \vartheta\bigl[s(\hat{A}_{\text{dn}}^{(w)}(E) - \hat{A}^{(\text{regular})}_{\text{dn}}(E))\bigr]\;
                                        \int_{T_{\text{min}}}^\infty \diff{T} \left\{
                                                                                    \sDelta\frac{\diff\sigma(E,T)}{\diff{T}} P_{\text{day}}(E) +
                                                                                    \frac{\diff\sigma_{\nu_x}(E,T)}{\diff{T}}
                                                                                \right\}
                        },
    \end{equation}
    which is just `high $-$ low', i.e. the difference between the average day-night asymmetries in the domains where it is expected
    to be higher than normal and where it is expected to be lower. In the above expression, $T_{\text{min}}$ is the threshold
    of recoil electron detection. A numerical evaluation of the above quantity, assuming $\hat{A}^{(\text{regular})}_{\text{dn}}(E)$
    to be the summer half-year day-night asymmetry and $T_{\text{min}} = 5~\text{MeV}$, amounts to
    \begin{equation}
        \Omega \approx 3.9\%,
    \end{equation}
    i.e. not less than the leading-order result. The above figure means that the number of events necessary to observe the effect
    is $\gtrsim 1 / \Omega^2 \sim 700$. Such a number of ${}^8$B neutrino events within a 2-month $\times$ 2-hour window
    is collected by a SuperK-size detector during two years (although, obviously, Super-Kamiokande does not provide the
    required energy resolution we are assuming here). Still, a typical next-generation detector will observe somewhat between
    Fig.~\ref{fig:Adn_E_midnightEffect} (perfect neutrino energy resolution) and Fig.~\ref{fig:Adn_T}
    (poor energy resolution, only recoil energies are observed), and, given larger detector volumes, additionally
    enhanced by the global fit mentioned above, the effect of the anomalous time-localized oscillatory
    deformation of the energy spectrum of the day-night effect will be quite observable.

    \section{Discussion and conclusion}\label{sec:Conclusion}
    In the previous sections, we have performed both an analytical and a numerical analyses of the possibility of observing the regeneration effect
    in the high-energy band of solar neutrinos (around 10~MeV) attempting to benefit as much as possible from quite predictable temporal (and seasonal)
    variations of the nighttime neutrino flux. It turned out that hidden effects in the nighttime oscillations can be revealed
    using quite simple techniques as the reduction of the observation term. This approach, however, requires certain
    neutrino energy resolution of the detector in order to make the effect more distinguishable. An improved angular resolution for recoil electrons
    in an elastic-scattering detector would substantially vivify the signature seen in the high-energy segment of Fig.~\ref{fig:Adn_T}.

    \vspace{0.5em}

    Observation of the peaks on the curves for the day-night asymmetry factors \eqref{Adn_exp_w} and \eqref{Adn_exp_w_theor}
    (Figs.~\ref{fig:Adn_T} and \ref{fig:Adn_E_midnightEffect}, respectively) gives
    an opportunity to precisely determine the oscillation parameters and/or the radii of the interfaces inside the Earth. Indeed,
    the peaks are very close to the maxima/minima of the oscillating function $\cos\bigl\{2\sDelta\psi_{n,j}(\chi + s \varepsilon;E) + s'(s-1) \frac{\pi}{4}\bigr\}$
    in Eq.~\eqref{Adn_analytical_w}, which relates the energies of the peaks to the radii of the interfaces or directly to $\sDelta{m^2}$
    (see Eqs.~\eqref{E_N_eqn} and \eqref{deltaE_N}). The more maxima one observes, the tighter the constraints on $\sDelta{m^2}$ one obtains.
    Note that, quite unexpectedly, the geographical placement of the detector (namely, its latitude) affects quite strongly the effects to be observed.

    Another feature of the effect in question we find important is that it is `global fit-ready', i.e., the data collected at
    different detectors can be subjected to a joint data processing procedure. Moreover, we would like to point out again that
    for the mixing parameters known to date \cite{PDG2014}, a detector placed in the Kamioka mine and a detector placed at
    the tropic will observe the contribution of the core-mantle interface counter-phase (see Fig.~\ref{fig:Adn_T}),
    which opens a possibility to isolate and magnify the anomalous contribution to the day-night asymmetry by
    subtracting the observational data at the two detectors. In principle though, even now, looking at the $A_{\text{dn}}(T)$
    plot in the latest report \cite{SuperK_DNA_latest} of the Super-Kamiokande collaboration on their day-night asymmetry observations,
    we cannot help observing a dip of the fit around $T = 12~\text{MeV}$ and an upheaval at $T = 14-16~\text{MeV}$
    that are similar to those presented in Fig.~\ref{fig:Adn_T}. Thus, it sounds particularly interesting to apply the temporal weighting
    to the Super-Kamiokande data collected to date and see if this will be able to reveal statistically-significant
    signatures of the time localization effect we are discussing here.

    \vspace{0.5em}

    It is worth mentioning that the time localization of the regeneration effect of solar neutrinos is stable with respect to
    perturbations of the radii of the interfaces inside the Earth within much less than one oscillation length. It is also stable
    with respect to the account of the non-pointlike neutrino source.
    In particular, non-sphericality of the Earth introduces minuscule effects for ${}^8$B neutrinos
    ($\ell_{\text{osc}} \sim 300~\text{km}$ at $E = 10~\text{MeV}$), in contrast to the case of ${}^7$Be neutrinos \cite{Smirnov2015}.
    The profiles of the transition zones between layers of different densities that are approximated by jump discontinuities
    within the PREM model do not affect noticeably our estimations, i.e. the positions and the amplitudes of the peaks
    on the $\hat{A}_{\text{dn}}^{(w)}(E)$ curve. The widths of these transition zones can be extended to tens of
    kilometres without changing the results. At the same time, a 300~km-wide interface between the outer core and the
    mantle will strongly suppress its contribution producing the lion's share of the effects discussed above.

    Apart from the density discontinuities mentioned above, our effect is sensible to the regions where the
    density profile is adiabatic, via the phases \eqref{deltaPsi_nj_def}. The latter ones, in turn, determine
    the positions of the peaks in Figs.~\ref{fig:Adn_E_midnightEffect} and \ref{fig:Adn_T} (see Eq.~\eqref{E_N_eqn}).
    A straightforward estimation shows that the effect of the matter shifts the phases $2\sDelta\psi_{n,j}$
    corresponding to the Earth's core by about $\pi$, while the leading-order account of the matter effect
    used in \eqref{E_N_LO} leads to an error in the phase about $\pi/6$. Thus, in principle, a redistribution
    of the Earth's density along its radius, say, inside the core, should lead to a more or less observable horizontal
    shift of the peaks at the energy spectra of the time-localized term in the day-night asymmetry.

    \vspace{0.5em}

    As we discussed in Sec.~\ref{sec:Localization}, the time-localized effects are quite robust with respect to the reduction
    of the observation term. That is, such a reduction (down to the time localization scale) improves the \emph{statistical} signal-to-noise
    ratio. Let us now assume that neutrino observations are characterized by \emph{systematic} errors as well, so that
    the observed mean values of $N_{\text{night,day}}^{(w)}$ differ from their predictions \eqref{Nnight}.
    It can be modelled as a redefinition of the expectations
    \begin{eqnarray}
        \mathds{E}[N_{\text{night,day}}^{(w)}] &=& \int\limits_{1~\text{year}}
                                                 \frac{\vartheta(\pm\pi/2 \mp \Theta_{\text{N}}(t)) w(t)\diff{t}}{0.5\text{~year}}
                                                 \int \Phi(E)\diff{E}\;
        \\
                            &&
                            \qquad\qquad\qquad \times \bigl\{ \sigma_{\nu_e}(E) P_{\nu_e}(\Theta_{\text{N}}(t); E) + \sigma_{\nu_x}(E) P_{\nu_x}(\Theta_{\text{N}}(t); E)
                            + \chi(E,t) \bigr\},
    \end{eqnarray}
    where $\chi(E, t)$ is a quantity we cannot precisely evaluate which represents the systematic uncertainty. The systematic uncertainties
    that (almost) do not depend on time, such as systematics in the cross sections, in the background
    due to decays of radioactive nuclei inside the Earth, solar neutrino spectrum systematics, etc., lead to cumulative terms in
    the day-night asymmetry, i.e. they are unaffected by temporal weighting. Therefore, the improvement of the signal-to-noise ratio
    corresponding to such systematics is even higher than that for statistical uncertainties, being approximately inversely
    proportional to the observation time $T_{\text{obs}}$ (see Sec.~\ref{sec:Localization}). In contrast, time-dependent
    (or time-unpredictable) systematic uncertainties, such as those due to local Earth's density deviations from the PREM model and
    due to time variations of the solar neutrino flux or spectrum due to helioseismic waves deep inside the Sun \cite{LENA},
    may, in principle, lead to time-localized contributions to the day-night asymmetry that
    will scale as $1/T_{\text{obs}}$, i.e. as the contributions (\emph{signal}) from the stationary points at the solstices.
    At the same time, interference of such effects with the time-localized day-night asymmetry discussed in the present paper
    might occur only if they happen to overlap with the observation window. Moreover, some of these
    short-duration effects, e.g., cosmic ray showers or certain anomalies in the solar neutrino flux,
    can be recognized at the detector and excluded from the data analysis. After all, most of them are very
    unlikely to produce oscillatory spectral signatures of the nighttime neutrino fluxes that are characteristic
    of the \emph{signal} $\hat{A}_{\text{dn}}^{(w)}(E)$. A clear exception is the effect of local density
    anomalies in the Earth's density profile right in the way of solar neutrinos within the observation window.
    For a window around the midnights close to the winter solstice which is of interest, such anomalies
    should lie close to the nadir. In principle, observation of such density anomalies, say, inside the inner core,
    could constitute a distinct experiment, though, to the best of our knowledge, the relevant effect should be at least several
    times smaller than the leading effect in Fig.~\ref{fig:Adn_E_midnightEffect} coming from a huge density discontinuity
    at the core-mantle interface.

    %

    \vspace{0.5em}

    In addition, we would like to point out another recipe for efficient observation of the day-night effect for
    ${}^8$B neutrinos, provided that one is in possession of a perfect energy-resolving neutrino detector (even though
    that does not seem feasible yet), i.e., one knows the energies for all neutrino events observed. In this case,
    it is much more efficient to observe the day-night effect through a `zone plate' or a `phase zone plate' \cite{ZonePlates},
    corresponding to the weighting functions of the form
    \begin{eqnarray}\label{w_zp}
        w &=& w_{\text{zp}}^\pm =
                    \begin{cases}
                          w_0 \vartheta\Bigl(\pm\cos2\sDelta\psi_{\text{out of the core}}(\Theta_{\text{N}}(t),E)
                                             \mp\cos2\sDelta\psi_{\text{into the core}}(\Theta_{\text{N}}(t),E) \Bigr), &
                          \Theta_{\text{N}}(t) < \Theta_{\text{N}}^{\text{core}},\\
                          0, & \Theta_{\text{N}}(t) \ge \Theta_{\text{N}}^{\text{core}},
                    \end{cases} \\
        \label{w_pp}
        w &=& w_{\text{pp}}^\pm =
                    \begin{cases}
                          w_0 \tilde\vartheta\Bigl(\pm\cos2\sDelta\psi_{\text{out of the core}}(\Theta_{\text{N}}(t),E)
                                             \mp\cos2\sDelta\psi_{\text{into the core}}(\Theta_{\text{N}}(t),E) \Bigr), &
                          \Theta_{\text{N}}(t) < \Theta_{\text{N}}^{\text{core}},\\
                          0, & \Theta_{\text{N}}(t) \ge \Theta_{\text{N}}^{\text{core}},
                    \end{cases}
    \end{eqnarray}
    where $\Theta_{\text{N}}^{\text{core}} \equiv \arcsin(r_{\text{core}} / r_{\text{Earth}})$ is the maximum nadir angle at which
    the neutrino crosses the core-mantle interface with $r = r_{\text{core}}$ and the two phase incursions $\sDelta\psi$
    are calculated from the two crossing points to the detector; $\vartheta(x)$ is the Heaviside step function and
    $\tilde\vartheta(x) \equiv |x|\, \vartheta(x)$. Note that here, the parameters of the weighting function are
    $\sDelta{m^2} / E$ and the radius of the core-mantle interface $r_{\text{core}}$. In contrast to the class of weighting functions
    studied so far, functions \eqref{w_zp}, \eqref{w_pp} depend on energies $E$ of individual neutrinos. Introduction of such
    weighting functions is equivalent to a modulation of the solar exposure function of the form
    \begin{eqnarray}
        \epsilon(\Theta) &\to& \const\times\epsilon(\Theta) \vartheta\Bigl(\pm\cos2\sDelta\psi_{\text{out of the core}}(\Theta,E)
                                              \mp\cos2\sDelta\psi_{\text{into the core}}(\Theta,E) \Bigr) \quad \text{(zone plate)},\\
        \epsilon(\Theta) &\to& \const\times\epsilon(\Theta) \tilde\vartheta\Bigl(\pm\cos2\sDelta\psi_{\text{out of the core}}(\Theta,E)
                                              \mp\cos2\sDelta\psi_{\text{into the core}}(\Theta,E) \Bigr) \quad\text{(phase zone plate)},
    \end{eqnarray}
    where the constants are defined by the normalization \eqref{expFunc_norm}.
    The purpose of the pairs of functions $w^\pm_{\text{zp,pp}}$ is a comparison of the day-night asymmetry weighted
    with $w_{\text{zp,pp}}^+$ with the one weighted with $w_{\text{zp,pp}}^-$,
    i.e., of the asymmetries observed when the core-mantle density jump makes a positive or a negative contribution to the nighttime
    neutrino observation probability $P_{\text{night}}(\Theta_{\text{N}}(t), E)$ (see Eq.~\eqref{valleyCliff}).

    \begin{figure}[tbh]
        \includegraphics[width=12cm]{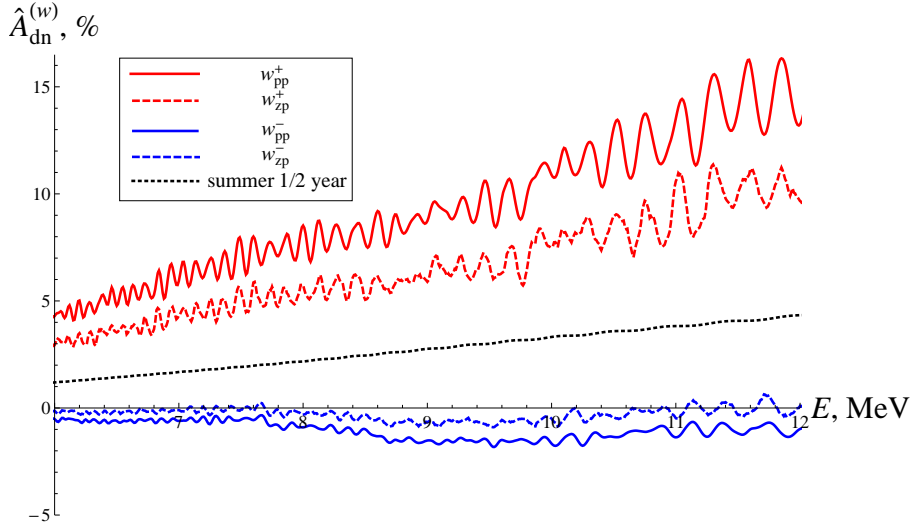}
        \caption{Day-night asymmetry factor $\hat{A}_{\text{dn}}^{(w)}(E)$ for a tropical detector ($\chi = 23.4\degree$), weighted
                 using a `phase zone plate' $w_{\text{pp}}^\pm$ (solid red/blue curves) and a `zone plate' $w_{\text{zp}}^\pm$
                 (dashed red/blue curves), see Eqs.~\eqref{w_zp},~\eqref{w_pp}. The dotted black curve shows the summer half-year day-night asymmetry
                 (solid red curve in Fig.~\ref{fig:Adn_E}).}
        \label{fig:Adn_E_zonePlate}
    \end{figure}

    The effect of using (phase) zone plates is presented in Fig.~\ref{fig:Adn_E_zonePlate},
    in terms of the weighted asymmetry factor $\hat{A}_{\text{dn}}^{(w)}$. It is clear that using zone-plate weighting
    is able to `focus' neutrinos that retain signatures of the Earth's core, using the positive interference of the expected neutrino
    flow and the weighting function, much like real, optical zone plates focus electromagnetic waves by rearranging the
    interference pattern. The order of the maximum of interference in our case is
    $\sim (r_{\text{Earth}} \pm r_{\text{core}}) / \ell_{\text{osc}} \sim 10-30$,
    thus, in order to apply the interference method of magnifying the contribution of the Earth's core, the detector should have at least a
    $3-10\%$ neutrino energy resolution. As a reward for a challenging task to achieve such a resolution, one would get a threefold-amplified
    effect of the core, i.e., for example, a $3^2 \times (T_{\text{core}} / T_{\text{night}}) \approx 2$-times
    reduced observation term needed to achieve the same level of statistical uncertainty (see Fig.~\ref{fig:Tcore_chi}).
    Note that the zone-plate observation can be combined with a full-year (non-weighted) observation of the cumulative contribution
    to the day-night asymmetry, thus reducing the statistical errors even further.

    \vspace{1em}

    The techniques outlined above will probably become important when next-generation detectors begin to operate.
    Still, from the numerical estimations presented in our paper, we are able to conclude that even now, the designers and technologists
    working on these future facilities should be aware of the observational opportunities
    that open if one adapts the experimental setup to the known properties of the neutrino sources and fluxes.
    Among the properties that turn out to be observationally useful, we have identified the time variations of the regeneration effect of
    solar neutrinos in the Earth, depending on the more or less known structure of the Earth's interior.

    \vspace{0.5em}

    After all, the effect we have studied in the present paper sounds quite peculiar, almost miraculous indeed~---
    the flux of solar neutrinos observed during the year has a point-like contribution coming from
    the midnights around Christmas. Even more, given the augmentation of the localized contribution
    in the tropical latitudes, we could figuratively say that if the biblical Magi had carried out
    neutrino observations along with astronomical ones usually attributed to them, then, analyzing the observational
    data, they could have observed the anomaly of the nighttime neutrino flux about a month before the
    birth of Jesus Christ (see the localization scales in Tables~\ref{tab:locScale}, \ref{tab:locScale_Tropic}) and
    could have arrived just in time in Bethlehem. Such an interpretation also reconciles the
    Gregorian and the Julian Christmases since both fit well within the localization window around the winter solstice.

    \vspace{0.5em}
    Leaving aside the discussion of possible religious parallels, we hope that the physical effect we have studied here
    will help observe solar neutrinos and efficiently process the relevant experimental data in the nearest future.

    \subsection*{Acknowledgments}
    The author is grateful to A.~E.~Lobanov and Y.~F.~Li for fruitful discussions and also to A.~E.~Lobanov
    for his kind advice on the structure of the manuscript. The numerical simulations described in the paper
    have been performed using the Supercomputing Cluster ``Lomonosov'' of the Moscow State University \cite{Lomonosov}.

\end{document}